\documentstyle[epsfig]{mn}

\input epsf
\input rotate

\newif\ifAMStwofonts

\ifoldfss

  \ifCUPmtlplainloaded \else

    \NewTextAlphabet{textbfit} {cmbxti10} {}

    \NewTextAlphabet{textbfss} {cmssbx10} {}

    \NewMathAlphabet{mathbfit} {cmbxti10} {} 

    \NewMathAlphabet{mathbfss} {cmssbx10} {} 

  \fi

  \ifAMStwofonts

    \ifCUPmtlplainloaded \else

      \NewSymbolFont{upmath} {eurm10}

      \NewSymbolFont{AMSa} {msam10}

      \NewMathSymbol{\upi}     {0}{upmath}{19}

      \NewMathSymbol{\umu}     {0}{upmath}{16}

      \NewMathSymbol{\upartial}{0}{upmath}{40}

      \NewMathSymbol{\leqslant}{3}{AMSa}{36}

      \NewMathSymbol{\geqslant}{3}{AMSa}{3E}

      \let\leq=\leqslant \let\le=\leqslant

      \let\geq=\geqslant \let\ge=\geqslant

    \fi

  \fi

\fi 

\ifnfssone

  \newmathalphabet{\mathit}

  \addtoversion{normal}{\mathit}{cmr}{m}{it}

  \addtoversion{bold}{\mathit}{cmr}{bx}{it}

  \newmathalphabet{\mathbfit} 

  \addtoversion{normal}{\mathbfit}{cmr}{bx}{it}

  \addtoversion{bold}{\mathbfit}{cmr}{bx}{it}

  \newmathalphabet{\mathbfss} 

  \addtoversion{normal}{\mathbfss}{cmss}{bx}{n}

  \addtoversion{bold}{\mathbfss}{cmss}{bx}{n}

  \ifAMStwofonts

    \ifCUPmtlplainloaded \else

      \UseAMStwoboldmath

      \makeatletter

      \new@mathgroup\upmath@group

      \define@mathgroup\mv@normal\upmath@group{eur}{m}{n}

      \define@mathgroup\mv@bold\upmath@group{eur}{b}{n}

      \edef\UPM{\hexnumber\upmath@group}

      \new@mathgroup\amsa@group

      \define@mathgroup\mv@normal\amsa@group{msa}{m}{n}

      \define@mathgroup\mv@bold\amsa@group{msa}{m}{n}

      \edef\AMSa{\hexnumber\amsa@group}

      \makeatother

      \mathchardef\upi="0\UPM19

      \mathchardef\umu="0\UPM16

      \mathchardef\upartial="0\UPM40

      \mathchardef\leqslant="3\AMSa36

      \mathchardef\geqslant="3\AMSa3E

      \let\leq=\leqslant \let\le=\leqslant

      \let\geq=\geqslant \let\ge=\geqslant

    \fi

  \fi

\fi 

\ifnfsstwo

  \DeclareMathAlphabet{\mathbfit}{OT1}{cmr}{bx}{it}

  \SetMathAlphabet\mathbfit{bold}{OT1}{cmr}{bx}{it}

  \DeclareMathAlphabet{\mathbfss}{OT1}{cmss}{bx}{n}

  \SetMathAlphabet\mathbfss{bold}{OT1}{cmss}{bx}{n}

  \ifAMStwofonts

    \ifCUPmtlplainloaded \else

      \DeclareSymbolFont{UPM}{U}{eur}{m}{n}

      \SetSymbolFont{UPM}{bold}{U}{eur}{b}{n}

      \DeclareSymbolFont{AMSa}{U}{msa}{m}{n}

      \DeclareMathSymbol{\upi}{0}{UPM}{"19}

      \DeclareMathSymbol{\umu}{0}{UPM}{"16}

      \DeclareMathSymbol{\upartial}{0}{UPM}{"40}

      \DeclareMathSymbol{\leqslant}{3}{AMSa}{"36}

      \DeclareMathSymbol{\geqslant}{3}{AMSa}{"3E}

      \let\leq=\leqslant \let\le=\leqslant

      \let\geq=\geqslant \let\ge=\geqslant

    \fi

  \fi

\fi 

\ifCUPmtlplainloaded \else

  \ifAMStwofonts \else 

    \def\upi{\pi}

    \def\umu{\mu}

    \def\upartial{\partial}

  \fi

\fi

\begin{document}

\title{Three-component St\"ackel potentials satisfying recent estimates of Milky Way parameters }

\author[B.Famaey \& H.Dejonghe]{B. Famaey$^{1}$ and H. Dejonghe$^{2}$\\
$^{1}$Institut d'Astronomie et d'Astrophysique CP226, Universit\'e Libre de Bruxelles, Boulevard du Triomphe, B-1050 Bruxelles, Belgium.\\
 Ph.D. student F.R.I.A., E-mail: bfamaey@astro.ulb.ac.be \\
$^{2}$Sterrenkundig Observatorium, Universiteit Gent, Krijgslaan 281, 
B-9000 Gent, Belgium
}

\date{Accepted ...
      Received 2002 july}

\pagerange{\pageref{firstpage}--\pageref{lastpage}}

\pubyear{2002}

\maketitle

\label{firstpage}

\begin{abstract}
We present a set of three-component St\"ackel potentials defined by five parameters and designed to model the Milky Way. We review the fundamental constraints that any model of the Milky Way must satisfy, including the most recent ones derived from Hipparcos data, and we study how the parameters of the presented potentials can vary in order to match these constraints. Five different valid potentials are presented and analyzed in detail: they are designed to be confronted with kinematical surveys in the future, by the construction of three-integral analytic distribution functions.
\end{abstract}


\section{Introduction}

The determination of the mass distribution and dynamical structure of the Milky Way is one of the fundamental tasks of Galactic Astronomy. Even though the Milky Way is a spiral barred galaxy, axisymmetric models are a necessary starting point for perturbation analysis and are thus a prerequisite if one wants to understand the bar on a theoretical basis. Caldwell \& Ostriker\ (1981), Rohlfs \& Kreitschmann\ (1988) and, recently, Dehnen \& Binney\ (1998) fitted axisymmetric mass models of the Milky Way to various measurements of the gravitational force field: they concluded that a wide variety of models can emerge from this fitting process and that the mass distribution of the Galaxy is still ill-determined.

In order to fully exploit kinematical stellar surveys, we should construct dynamical models based on Jeans\ (1915) theorem. This theorem states that the phase space distribution function of a stellar system in a steady state depends only on three isolating integrals of the motion: numerical experiments\ (Ollongren 1962; Innanen \& Papp 1977; Richstone 1982) showed that most orbits in realistic galactic potentials admit three such integrals. The third integral, in addition to the binding energy and the vertical component of the angular momentum, is  not analytic in a general potential: it is possible to define an approximate third integral specific to particular orbital families\ (de Zeeuw, Evans \& Schwarzschild 1996; Evans, H\"afner \& de Zeeuw 1997; De Bruyne, Leeuwin \& Dejonghe 2000) or to foliate phase space with tori on which three action integrals can be defined in a potential that differs from the non-integrable one by only a small amount\ (Kaasalainen \& Binney 1994; Binney 2002). Instead, we choose to construct models with an exact analytic third integral by using St\"ackel potentials\ (St\"ackel 1890). These potentials were introduced into stellar dynamics by Eddington\ (1915) and have since been used in a number of papers\ (e.g. Lynden-Bell 1962; de Zeeuw 1985; Dejonghe 1993; Sevenster, Dejonghe \& Habing 1995; Durand, Dejonghe \& Acker 1996; Bienaym\'e 1999) : in fact, the regularity of typical galactic potentials can be understood in terms of their proximity to St\"ackel potentials\ (e.g. Gerhard 1985).

Our long term goal is to constrain the mass distribution of the Galaxy by using kinematical stellar surveys: we shall choose a St\"ackel potential, and use the quadratic programming technique described by Dejonghe\ (1989) to determine, for the available surveys, distribution functions in the space of the integrals of the motion\ (see Famaey, Van Caelenberg \& Dejonghe 2002). Then we shall compare the resulting predictions with the surveys. The potential will be modified in the light of that comparison. Thus the first step is to show that different St\"ackel potentials are compatible with the standard constraints for a mass model of the Milky Way.

In this paper, our goal is to show that a wide variety of simple St\"ackel potentials can fit most known parameters of the Milky Way (including Hipparcos latest findings). In order to do this, we continue the work of Batsleer \& Dejonghe\ (1994, hereafter BD) who presented a set of simple St\"ackel potentials with two mass components (halo and disc) and a flat rotation curve, that we generalize by adding a thick disc to them since its existence as a separate stellar component is now well documented\ (Ojha et al. 1994; Chen et al. 2001). These new potentials are described by five parameters and we will show that many different combinations of these parameters are consistent with fundamental constraints for a mass model of the Milky Way.

In the next section, we review the observational constraints that any mass model of the Milky Way must satisfy. In section 3, we present the mathematical form of the set of St\"ackel potentials studied in this paper, and we choose selection criteria in the light of the constraints reviewed in section 2. In section 4, we examine the consequences of those selection criteria and we present five potentials differing in terms of form and features, and that are plausible for the Milky Way. For the conclusions, we refer to section 5.

\section{Observational constraints}

Recently, Hipparcos data\ (ESA 1997) have brought nearly definitive answers to some long-standing questions in Galactic Astronomy. In this section, we review the determination of those galactic fundamental parameters that any Milky Way potential must match.

\subsection{Galactocentric radius of the sun}

The distance of the sun to the galactic center is difficult to estimate: the direct method is to compare the average radial velocities with the average proper motions of maser spots in star forming regions near the galactic center, but this method needs very accurate observations and it is affected by extinction. If the density of the objects of the stellar halo of the Milky Way peaks at the galactic center, the galactocentric radius of the sun can then also be measured by determining the distance of this density peak: the problem of this method is that any uncertainty in the absolute magnitude of the stellar candles will reverberate in the estimation of the galactocentric distance of the sun. The measurements of this distance have been reviewed by Reid\ (1993), and they tend to approach 8 kpc. In this paper, we assume this estimate of 8 kpc for the galactocentic solar radius, with an uncertainty of the order of 0.5 kpc.

\subsection{Flat rotation curve} 

The stars of the disc travel in nearly circular orbits around the galactic center. The determination of the rotation curve $v_{c}(\varpi)$, where $\varpi$ is the galactocentric radius, and in particular $v_{c}(\varpi_{\odot})$ is one of the hardest problems in Galactic Structure (see section 2.3 for the local shape of the rotation curve and the determination of $v_{c}(\varpi_{\odot})$). The rotation curve is determined by observations of the kinematics of the gas, and in particular of the neutral hydrogen 21 cm line, but the rotation curve is not well established for the galactic radii $\varpi > \varpi_\odot$. Observations of outer spiral galaxies indicate that the rotation curve remains more or less flat after attaining its maximum\ (e.g. Casertano \& van Gorkom 1991): the rotation curve of a mass model of the Milky Way therefore is likely to behave similarly.

\subsection{Oort constants and local circular speed}

As the global shape of the rotation curve is not known precisely, the determination of its local shape in the solar neighbourhood is fundamental for a better knowledge of local Galactic Structure. Lindblad\ (1925) and Oort\ (1927a,b) developed the model of differential axisymmetric rotation with $\Omega = \frac{v_{c}}{\varpi}$ depending only on the distance $\varpi$ to the galactic center. Oort\ (1927a) introduced two constants (the Oort constants $A$ and $B$), that can be determined from proper motions of neighbouring stars and that are directly related to the local shape of the rotation curve. Indeed, for a star of the solar neighbourhood on a circular orbit, Taylor expanding $\Omega(\varpi)$ to first order in $(\varpi-\varpi_{\odot})$ yields for the radial and transverse velocities (with $d$ denoting the distance to the sun): 

\begin{equation}
v_r = A \, d \, {\rm sin}2l
\end{equation}
\begin{equation}
v_t = A \, d \, {\rm cos}2l + B \, d
\label{eq:vt}
\end{equation}
with
\begin{equation}
     A = \frac{1}{2}(\frac{v_c}{\varpi}-\frac{{\rm d}v_c}{{\rm d}\varpi})_{\varpi_\odot} , B = -\frac{1}{2}(\frac{v_c}{\varpi}+\frac{{\rm d}v_c}{{\rm d}\varpi})_{\varpi_\odot}
\end{equation}

Kuijken \& Tremaine\ (1991) showed that the Taylor expansion terms arising from non-circularity of the orbits are negligible, just as they should be if the Milky Way is axisymmetric. Oort\ (1927b) showed for the first time this sinusoidal effect of the galactic rotation on the radial velocities and on the proper motions. He found $A = 19 {\rm km}\,{\rm s}^{-1}{\rm kpc}^{-1}$ and $B = -24 {\rm km}\,{\rm s}^{-1}{\rm kpc}^{-1}$. Indeed, by fitting Eq.(\ref{eq:vt}) to observed proper motions, one can determine $A$ and $B$ if one is sure that the frame is not rotating. This last requirement was pretty unsure before the Hipparcos mission. The most reliable determination of the Oort constants based on the proper motions of the Cepheids that were measured by Hipparcos has been derived by Feast \& Whitelock\ (1997) who found $A = 14.82 \pm 0.84 {\rm km}\,{\rm s}^{-1}{\rm kpc}^{-1}$ and $B = -12.37 \pm 0.64 {\rm km}\,{\rm s}^{-1}{\rm kpc}^{-1}$. These values were confirmed by Mignard\ (2000) who found $A = 14.5 \pm 1.0 {\rm km}\,{\rm s}^{-1}{\rm kpc}^{-1}$ and $B = -11.5 \pm 1.0 {\rm km}\,{\rm s}^{-1}{\rm kpc}^{-1}$ by using proper motions of distant giants. This indicates that the rotation curve is slightly declining in the solar neighbourhood. 

The determination of $v_{c}(\varpi_{\odot})$ follows from the determination of the Oort constants: using the values of Feast \& Whitelock\ (1997), we find $v_{c}(\varpi_{\odot})=(218 \pm 8{\rm km}\,{\rm s}^{-1})(\varpi_{\odot}/8{\rm kpc})$, a value which is consistent with $v_{c}(\varpi_{\odot})= 220 \, {\rm km}\,{\rm s}^{-1}$ that we choose to adopt in this paper. Since we impose $v_{c}(\varpi_{\odot})= 220 \, {\rm km}\,{\rm s}^{-1}$ and $\varpi_\odot = 8 \pm 0.5 \, {\rm kpc}$ for all the potentials defined in section 3, the value of $A-B = \frac{v_{c}(\varpi_{\odot})}{\varpi_\odot}$ is in the fixed interval $27.6 \pm 1.7 {\rm km}\,{\rm s}^{-1}{\rm kpc}^{-1}$. This interval is in accordance with the one deduced from the values of $A$ and $B$ determined by Feast \& Whitelock\ (1997), i.e. $\frac{v_c(\varpi_\odot)}{\varpi_\odot} = 27.2 \pm 0.9{\rm km}\,{\rm s}^{-1}{\rm kpc}^{-1}$. So, the only relevant constraint relative to the Oort constants for our potentials will be the value of $\frac{{\rm d}v_c}{{\rm d}\varpi}(\varpi_\odot)$.

However, The Oort constants are not considered as a very strong constraint in this paper. Indeed, the measurement of the proper motion of the compact radio source Sgr A*\ (Backer 1996), seems to indicate that $A-B = 30.1 \pm 0.8 {\rm km}\,{\rm s}^{-1}{\rm kpc}^{-1}$ when assuming that Sgr A* is stationary with respect to the galactic center. This is inconsistent with the determination based on Hipparcos data, and leaves us with an uncertainty which still awards a resolution.

\subsection{Local dynamical mass}

The mass density in the solar neighbourhood $\rho_{\odot}$ is an essential constraint for any mass model. It can be surmised that this parameter is not directly observed: it is deduced from the positions and velocities of tracer stars in the direction perpendicular to the galactic plane (the $z$-direction), using the Boltzmann and Poisson equations in various forms.

Jeans equations (first order moments of the Boltzmann equation) imply that the vertical acceleration $K_{z}$ is related to the vertical velocity dispersion $\sigma_{z}^2$ and to the vertical number density $g(z)$ of a population of stars by:

\begin{equation}
K_{z}(z) = \sigma_{z}^2 \frac{{\rm d}ln(g(z)/g(0))}{{\rm d}z} ,
\end{equation}
if the vertical motion can be separated from the radial and azimuthal motion of the stars (Oort-Lindblad approximation), if the velocity ellipsoid is aligned with the cylindrical coordinate axes (i.e $\langle v_{\varpi}v_{z} \rangle = 0$), and if the population is isothermal (i.e. $\sigma_{z}^2$ is constant as a function of $z$).

Oort\ (1932; 1960) applied this now classical formula to late-type stars (A to M) with the assumption that they were old enough to have become dynamically well mixed in the $z$-direction. Then he derived $\rho_{\odot}$ using Poisson equation for nearly circular orbits:

\begin{equation}
4 \pi G \rho_{\odot} = -\frac{{\rm d}K_z}{{\rm d}z} -2(A^2-B^2)
\label{eq:Pois}
\end{equation}

Oort found $\rho_{\odot}=0.09 M_{\odot} {\rm pc}^{-3}$ in 1932 and $\rho_{\odot}=0.15 M_{\odot} {\rm pc}^{-3}$ in 1960: this last result indicated that there might be a lot of dark matter in the disc, by comparison with star counts.

Afterwards, many other similar studies\ (Yasuda 1961; Eelsalu 1961; Woolley \& Stewart 1967; Turon Lacarrieu 1971; Gould \& Vandervoort 1972; Jones 1972; Balakirev 1976; Hill, Hilditch \& Barnes 1979) gave discrepant results, suffering from inhomogeneities in the data, systematic errors due to the use of photometric distances, and undersampling near the galactic plane. Because of this undersampling, some studies were even based on young O and B stars, assuming that the gas and dust out of which they recently formed was already roughly relaxed\ (Stothers \& Tech 1964; von Hoerner 1966).

More recently, Bahcall\ (1984a,b,c) described the disc matter as a series of isothermal components and analyzed the nonlinear self-consistent equations in which the matter produces the potential (Poisson equation) and is also affected by the potential (Jeans equations for each isothermal component). Bahcall, Flynn \& Gould\ (1992) applied this method to a sample of K giants and found $\rho_{\odot}=0.26 M_{\odot} {\rm pc}^{-3}$, a result leading once again to the presence of dark matter in the disc.

At about the same time, Kuijken \& Gilmore\ (1989a,b,c, 1991) used another method which does not use the Jeans equations, inspired by the method of von Hoerner\ (1960). This method is based on the assumption that the phase space distribution function $F_z(z,v_z)$ of any tracer population depends only on
\begin{equation}
E_z = V_z(z) + \frac{1}{2} v_z^2
\end{equation}
where
\begin{equation}
- \frac{{\rm d}V_z(z)}{{\rm d}z} = K_z(z).
\end{equation}

\noindent
This assumption follows from the classical separability of the vertical motion of the stars and from the Jeans\ (1915) theorem in one dimension. The density in configuration space $g(z)$ of  a tracer population is then related to its density in phase space by the integral equation:
\[g(z) = \int_{-\infty}^{\infty} F_z(z,v_z) {\rm d}v_z = 2 \int_{V_z}^{\infty} \frac{F_z(E_z)}{\sqrt{2(E_z - V_z)}} {\rm d}E_z \]
\begin{equation}
=f(V_z)=f(V_z(z))
\label{eq:int}
\end{equation}

\noindent
So, there is a unique relation between $g(z)$ and $F_z(E_z)$: Kuijken \& Gilmore\ (1989c, 1991) inverted the Abel transform (\ref{eq:int}) and used the space density profile $g(z)$ of distant K stars to predict their velocity distribution at different heights for different $V_z(z)$. Then they compared these predictions to the velocity data and used a maximum likelihood technique to select the best-fitting potential. The data they used were too far from the plane to constrain $\rho_{\odot}$ and they found a surface mass density between $z=\pm 1.1$ kpc of  $71 M_{\odot} {\rm pc}^{-2}$, a result rejecting the presence of dark matter in the disc.

So, all the investigations between 1932 and the Hipparcos mission had failed to converge to a reliable determination of $\rho_{\odot}$: they were all very uncertain, essentially because of inhomogeneities in the tracer samples, undersampling near the equatorial plane and the use of photometric distances. Hipparcos data solved all these problems: the completeness of the Hipparcos survey for stars brighter than $m_v = 8.0$ solved the homogeneity problem, the dense probe near the plane eliminated the undersampling, while the accurate parallaxes eliminated the use of photometric distances.

Cr\'ez\'e et al.\ (1998) and Holmberg \& Flynn\ (2000) used Eq. (\ref{eq:int}) to determine $\rho_{\odot}$ from complete samples of nearby A-F stars. Given the functions $g(z)$ and $F_z(z=0,v_{z=0}^2) = F_z(E_z)$ from the observed vertical density and velocity distribution, the function $V_z(z)$ can be derived. Then, to determine $\rho_{\odot}$, the Oort constants have to be used in the Poisson equation (\ref{eq:Pois}): these are also much better known since the Hipparcos mission (see section 2.3). Cr\'ez\'e et al.\ (1998) estimated $\rho_{\odot} = 0.076 \pm 0.015 M_{\odot} {\rm pc}^{-3}$ while Holmberg \& Flynn\ (2000) estimated $\rho_{\odot} = 0.102 \pm 0.010 M_{\odot} {\rm pc}^{-3}$. These differences between local density estimates using almost the same Hipparcos data are due to different a priori hypothesis on the vertical potential $V_z$. Recent investigations at Strasbourg University using a local Hipparcos sample combined with two samples at the galactic poles confirm the values obtained by Holmberg \& Flynn\ (2000) (Siebert, Bienaym\'e \& Soubiran 2002).

We conclude that any mass model of the Milky Way must have its solar neighbourhood mass density in the range $0.06 M_\odot {\rm pc}^{-3} \leq \rho_{\odot} \leq 0.12 M_\odot {\rm pc}^{-3}$ in order to match Hipparcos latest findings (rejecting the presence of a disc-like dark matter component). However, it should be remarked that all the results presented here are based on the Oort-Lindblad approximation, which may not be valid: therefore we do not exclude that another subset of potentials, that do not match our imposed constraints based on Oort-Lindblad approximation, could also be useful for dynamical modeling of the Milky Way.

 \section{The Potentials}

The St\"ackel potentials are non-rotating potentials for which the Hamilton-Jacobi equation is separable and for which all orbits admit three analytic integrals of the motion\ (St\"ackel 1890): they form the most general set of potentials that contain one free function, for which three exact integrals of the motion are known, and which can be relevant as models for a global potential in galactic dynamics\ (Lynden-Bell 1962). In this section, we present the mathematical form of the class of St\"ackel potentials that are studied in this paper.

\subsection{Coordinate system}

Axisymmetric St\"ackel potentials are best expressed in spheroidal coordinates $(\lambda,\phi,\nu)$, with $\lambda$ and $\nu$ the roots for $\tau$ of
\begin{eqnarray}
        \frac{\varpi^2}{\tau+\alpha}+\frac{z^2}{\tau+\gamma}=1 & & \alpha < \gamma < 0,
\end{eqnarray}
and $(\varpi,\phi,z)$ cylindrical coordinates. The parameters $\alpha$
and $\gamma$ are both constant and we assume them smaller than zero. It is convenient to define the axis ratio of the coordinate surfaces as $\epsilon = \frac{a}{c}$ with $\alpha = -a^2$ and $\gamma = -c^2$. Together with the focal distance $\Delta = \sqrt{\gamma - \alpha}$, the axis ratio defines the coordinate system.

\subsection{Three-component St\"ackel potentials}

An axisymmetric potential is of St\"ackel form if there exists a spheroidal
coordinate system $(\lambda,\phi,\nu)$ in which the potential can be
written as
\begin{equation}
        V(\lambda,\nu) = - \frac{f(\lambda) - f(\nu)}{\lambda - \nu},
\end{equation}
for an arbitrary function $f(\tau) = (\tau+\gamma)G(\tau)$, $G \geq 0$, $\tau =
\lambda,\nu$. The function $-G(\lambda)$ then represents the potential
in the $z=0$ plane.

The Milky Way is composed of several mass components: the bulge, the thin disc, the thick disc, the stellar halo, the dark halo, and the (dynamically insignificant) interstellar medium. It is not fundamental that a mass model aknowledges explicitly the existence of each of these components. For example, BD presented a set of two-component (halo-disc) St\"ackel potentials with a flat rotation curve.

Our goal is to show that different St\"ackel potentials are able to fit the latest estimates for the fundamental parameters of the Galaxy. We first generalize the potentials of BD by adding a thick disc to them since its existence as a separate stellar component is now well established\ (Ojha et al. 1994; Chen et al. 2001). Our potentials have thus three mass components: two ``flat'' components and one spheroidal. The spheroidal component accounts for the stellar and dark halo, and we shall see in section 4 that our potentials turn out to have an effective bulge, which enables us to avoid the explicit introduction of a bulge-component.

We assume that all three components of our potentials generate a St\"ackel potential, with three different coordinate systems but the same focal distance. It is straightforward to show that the superposition of three St\"ackel potentials is still a St\"ackel potential when all three coordinate systems have the same focal distance. Although the functions $f_{\rm thin}$, $f_{\rm thick}$ and $f_{\rm halo}$ are arbitrary, we assume that they each generate a Kuzmin-Kutuzov potential, defined by 
\begin{equation}
G(\tau) = \frac{GM}{\sqrt{\tau}+c}
\label{eq:KK}
\end{equation}
with $M$ the total mass of the system. Such a potential becomes a point mass potential ($V=-\frac{GM}{\varpi}$) in the Galactic Plane when $\lambda \rightarrow \infty$. We use this potential essentially because it is an extremely simple but representative St\"ackel potential: we will show that it is not necessary to use complicated St\"ackel potentials in order to match all the known and most recently determined parameters of the Milky Way.

Near the center, in a meridional plane, the lines of constant mass density corresponding to a St\"ackel potential are approximately ellipsoidal\ (de Zeeuw, Peletier \& Franx 1986). For a Kuzmin-Kutuzov potential\ (see e.g. Dejonghe \& de Zeeuw 1988), when $a>c$, the isodensity surfaces are flattened oblate spheroids, and increasing $\epsilon=\frac{a}{c}$ produces more flattening. So, the ratio $\epsilon$ has to be high for the thin disc, intermediate for the thick disc and close to unity for the halo.

We first define a class of dimensionless potentials $V_p$ in dimensionless units $(\varpi_p,z_p)$, with a focal distance $\Delta=1$ for all three coordinate systems and the central value of the potentials equal to $-1$. Each of these potentials is a superposition of three Kuzmin-Kutuzov potentials in three different coordinate systems:
\begin{eqnarray}
\lefteqn{V_p(\lambda_{\rm thin},\lambda_{\rm thick},\lambda_{\rm halo},\nu_{\rm thin},
     \nu_{\rm thick},\nu_{\rm halo}) = \nonumber} \\
\lefteqn{-k_{\rm thin}\frac{f_{\rm thin}(\lambda_{\rm thin})-f_{\rm thin}(\nu_{\rm thin})}{\lambda_{\rm thin}-\nu_{\rm thin}} \nonumber} \\
\lefteqn{-k_{\rm thick}\frac{f_{\rm thick}(\lambda_{\rm thick})-f_{\rm thick}(\nu_{\rm thick})}{\lambda_{\rm thick}-\nu_{\rm thick}} \nonumber} \\
\lefteqn{-(1-k_{\rm thin}-k_{\rm thick})\frac{f_{\rm halo}(\lambda_{\rm halo})-f_{\rm halo}(\nu_{\rm halo})}{\lambda_{\rm halo}-\nu_{\rm halo}}}
\end{eqnarray}
This new class of potentials is thus defined by five parameters (the three axis ratios of the coordinate surfaces and the relative contribution of the thin and thick disc masses to the total mass, i.e. $k_{\rm thin}$ and $k_{\rm thick}$), which is a reasonable augmentation with respect to the BD potentials that were defined by three parameters.

We denote
\begin{eqnarray}
\lefteqn{\alpha_{\rm thin}-\alpha_{\rm thick} = \gamma_{\rm thin}-\gamma_{\rm thick} = q_1 \ge 0 \nonumber} \\
\lefteqn{\alpha_{\rm thin}-\alpha_{\rm halo} = \gamma_{\rm thin}-\gamma_{\rm halo} = q_2 \ge q_1 \ge 0.}
\end{eqnarray}
So, we can express the class of potentials $V_p$ as a function of $\lambda_{\rm thin}$, $\nu_{\rm thin}$ and the two constants $q_1$ and $q_2$ (we also use Eq. (\ref{eq:KK})) to give the final form of $V_p$:
\begin{eqnarray}
\lefteqn{V_p(\lambda_{\rm thin},\nu_{\rm thin}) = - GM (\frac{k_{\rm thin}}{\sqrt{\lambda_{\rm thin}} + \sqrt{\nu_{\rm thin}}} \nonumber} \\
\lefteqn{+ \frac{k_{\rm thick}}{\sqrt{\lambda_{\rm thin}+q_1} + \sqrt{\nu_{\rm thin}+q_1}} \nonumber} \\
\lefteqn{+ \frac{1-k_{\rm thin}-k_{\rm thick}}{\sqrt{\lambda_{\rm thin}+q_2} + \sqrt{\nu_{\rm thin}+q_2}}) } 
\end{eqnarray}

The dimensionless rotation curve corresponding to such potentials is given by:
\begin{eqnarray}
\lefteqn{v_c^2(\varpi_p) = \varpi_p \partial_{\varpi_p}V_p(\varpi_p,0) \nonumber} \\
\lefteqn{= GM\varpi_p^2(\frac{k_{\rm thin}}{\sqrt{\lambda_{\rm thin}}(\sqrt{\lambda_{\rm thin}}+c_{\rm thin})^2} \nonumber} \\
\lefteqn{\frac{k_{\rm thick}}{\sqrt{\lambda_{\rm thin}+q_1}(\sqrt{\lambda_{\rm thin}+q_1}+\sqrt{c_{\rm thin}^2+q_1})^2} \nonumber} \\
\lefteqn{\frac{1-k_{\rm thin}-k_{\rm thick}}{\sqrt{\lambda_{\rm thin}+q_2}(\sqrt{\lambda_{\rm thin}+q_2}+\sqrt{c_{\rm thin}^2+q_2})^2} )} 
\end{eqnarray}
where $\varpi_p$ denotes the dimensionless galactocentric radius.
We shall impose constraints on the shape and flatness of the rotation curve in section 3.3.

In order to transform these dimensionless potentials into dimensional ones for the Milky Way, we denote the dimensionless radius where the rotation curve attains its first maximum as $\varpi_{p,M}$: since the Milky Way attains its global amplitude of 220 km/s for the first time at a radius of about 1.5 kpc\ (Fich \& Tremaine 1991), we define a distance scale factor $r_S=\frac{1.5{\rm kpc}}{\varpi_{p,M}}$. The conversion between dimensionless and dimensional distances is then given by:
\begin{equation}
                    \left\{ 
                        \begin{array}{l}
                            \varpi ({\rm kpc})= r_S \varpi_p \\
                            z ({\rm kpc}) = r_S z_p  \\ 
                        \end{array}
                    \right.
\end{equation}

Then, the total mass $M$ of the Galaxy is adjusted in such a way that the dimensional circular velocity at the solar radius ($\varpi_\odot = 8 \pm 0.5 \, {\rm kpc}$, see section 2.1) is equal to $220\,{\rm km}\,{\rm s}^{-1}$: we obtain thus a minimum and a maximum value for $M$, for the two extreme values of the galactocentric distance of the sun. That adjustment also fixes the local mass density in the solar neighbourhood $\rho_\odot$ (see section 3.3).

\subsection{Selection criteria}

We shall now establish the features that a potential (as defined in section 3.2) must have to be considered as a plausible potential for the Milky Way, in the light of the observational constraints reviewed in section 2.

By definition of the dimensional potentials (see section 3.2), the local circular speed $v_c(\varpi_\odot)$ is equal to $220\,{\rm km}\,{\rm s}^{-1}$ for all the potentials. The first fundamental selection criterion is the flatness of the rotation curve: this feature can be examined in the dimensionless frame-work. Even the BD potentials, with only two mass components, could produce many different shapes of rotation curves. So, we adopt the same simple diagnostic as BD: we denote $\varpi_{p,M}$ the dimensionless radius where the circular speed attains its first maximum and we look for a range in $\varpi$ where $v_c(\varpi)$ remains larger than 80\% of the maximum velocity and is thus more or less constant. We denote $\varpi_{p,F}$ the dimensionless radius where $v_c(\varpi_{p,F})=0.8v_c(\varpi_{p,M})$. A rotation curve is considered sufficiently flat if:
\begin{equation}
E_F = \frac{\varpi_{p,F} - \varpi_{p,M}}{\varpi_{p,M}} > 8
\label{eq:ef}
\end{equation}
This is a minimum requirement.

The second selection criterion is based on the latest determinations of the Oort constants. For all our potentials,
\begin{equation}
\frac{v_c(\varpi_\odot)}{\varpi_\odot} = 27.6 \pm 1.7{\rm km}\,{\rm s}^{-1}{\rm kpc}^{-1}
\end{equation}
which is in accordance with the value determined by Feast \& Whitelock\ (1997), $\frac{v_c(\varpi_\odot)}{\varpi_\odot} = 27.2 \pm 0.9{\rm km}\,{\rm s}^{-1}{\rm kpc}^{-1}$. The first derivative of the circular velocity in the solar neighbourhood corresponding to the Oort constants found by Feast \& Whitelock\ (1997) is $\frac{{\rm d}v_c}{{\rm d}\varpi}(\varpi_\odot) = -2.4 \pm 1.2{\rm km}\,{\rm s}^{-1}{\rm kpc}^{-1}$. Our potentials have two extreme values for this derivative, depending on the position of the sun; in order to fit the above interval, we select the potentials such that:
\begin{equation}
                    \left\{ 
                        \begin{array}{l}
                            {\rm max}(\frac{{\rm d}v_c}{{\rm d}\varpi}(\varpi_\odot)) > -3.6{\rm km}\,{\rm s}^{-1}{\rm kpc}^{-1} \\
                            {\rm min}(\frac{{\rm d}v_c}{{\rm d}\varpi}(\varpi_\odot)) < -1.2{\rm km}\,{\rm s}^{-1}{\rm kpc}^{-1}  \\ 
                        \end{array}
                    \right.
\end{equation}
This feature is not as essential as the flatness of the rotation curve, because of the intriguing measurement of the proper motion of Sgr A* (see section 2.3).

The last selection criterion is the local mass density in the solar neighbourhood: this number is determined by the adopted total mass $M$. We look for its values in both extreme positions for the sun $(\varpi_\odot,z_\odot)=(7.5,0.004)\,{\rm kpc}$ and $(\varpi_\odot,z_\odot)=(8.5,0.02)\,{\rm kpc}$. Following the Hipparcos latest findings reviewed in section 2.4, we select the potentials such that:
\begin{equation}
                    \left\{ 
                        \begin{array}{l}
                            {\rm max}(\rho_\odot) > 0.06 M_\odot {\rm pc}^{-3} \\
                            {\rm min}(\rho_\odot) < 0.12 M_\odot {\rm pc}^{-3}  \\ 
                        \end{array}
                    \right.
\end{equation}
This feature is more important than the Oort constants: we give the priority to that constraint, contrarily to Sevenster et al.\ (2000) who studied the structure of the inner Galaxy, and therefore constructed St\"ackel potentials with reasonable values for the Oort constants but values for $\rho_\odot$ that are too low.

The total mass of the Galaxy, the mass fractions of the discs and the flattening and scale length of the components are not well established observationally and are not considered as fundamental constraints for the potential.

\begin{figure*}
\psfig{figure=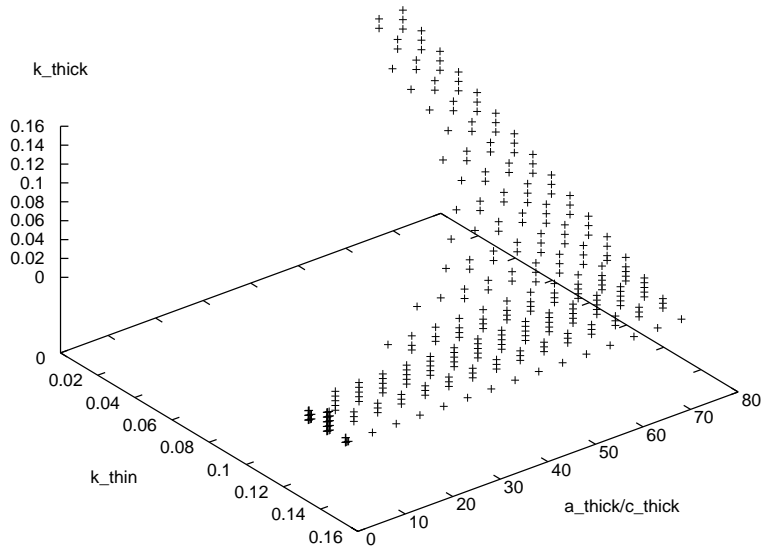,width=12.cm,angle=-90} \caption{This ``winding staircase'' figure displays the possible values of the coordinate axis ratio $\epsilon=\frac{a}{c}$ for the thick disc and the possible values of the contributions $k$ of the discs to the total mass in order to satisfy the selection criteria of section 3.3 (for fixed values $\epsilon_{\rm thin} = 75$ and $\epsilon_{\rm halo} = 1.02$, and with the thick disc always thicker than the thin disc, i.e. $\epsilon_{\rm thick} < 75$). It gives a rough vision of the region of parameter space that satisfies the criteria. Only the region $1.3 \le \epsilon_{\rm thick} \le 2$ is in fact relevant for a model of the Milky Way (see section 4.2).}
\end{figure*}

\section{Results}

Our goal is now to find and select, among the class of potentials defined in section 3, some representative potentials that differ with respect to their form and features, and that satisfy the fundamental constraints reviewed in section 2 and quantified in section 3.3. As it is difficult to visualize a five-dimensional parameter space, we shall first visualize its structure when $\epsilon_{\rm thin}$ and  $\epsilon_{\rm halo}$ are fixed (the ``winding staircase''). Then we shall restrict the parameter space by imposing additional constraints on the flatness of the thick disc.  We shall finally select five representative potentials, listed in Table 5. 

\subsection{The ``winding staircase''}

As an example of the consequences of the choice of the selection criteria of section 3.3, we look for all the values of $\epsilon_{\rm thick}$, $k_{\rm thin}$ and $k_{\rm thick}$ that yield potentials satisfying the selection criteria for $\epsilon_{\rm thin} = 75$ and $\epsilon_{\rm halo} = 1.02$ (with the thick disc always thicker than the thin disc, i.e. $\epsilon_{\rm thick} < 75$). The accordance with the selection criteria results from a precise mixing of the two discs and of the halo. Figure 1 illustrates the volume in parameter space corresponding to these satisfactory potentials: the volume looks like a ``winding staircaise''.  When $k_{\rm thick} = 0$, we see that all the values of $\epsilon_{\rm thick}$ are allowed when $13\% \le k_{\rm thin} \le 15\%$, which results from the fact that no thick disc is in fact present. When $k_{\rm thin}$ is close to its maximum possible value (15\%), $k_{\rm thick}$ has to be zero except when $\epsilon_{\rm thick}$ is very close to 1, i.e. when the thick disc is a pretty round component (similar to the halo) and helps to keep the rotation curve flat. When $k_{\rm thin}$ is smaller than 15\%, the possibilities for $k_{\rm thick}$ are more numerous, i.e. the thick disc can take a part of the mass. When $k_{\rm thin}$ attains the critical value of 12\%, the volume is inflected and the thick disc has to be thin and non-zero in order to counter-balance the lack of mass in the thin disc. Decreasing the mass of the thin disc forces the thick disc to become thinner and more massive, yielding the ``winding staircase'' volume of Figure 1. The similar volumes in parameter space for $\epsilon_{\rm thin} > 75$ have the same form, and are bigger essentially because there is more freedom for $\epsilon_{\rm thick}$.

\subsection{Constraints on the scale height of the thick disc}

In Figure 1, there are some solutions with a thick disc more massive than the thin disc. Even though the mass fractions of the discs and the flattening of the components are not well established and should be tested in a dynamical study, we know that the mass fraction of the thick disc is smaller than that of the thin disc (and represent at most 13\% of the local thin disc density in the solar neighbourhood) and the latest determination of the thick disc scale height based on star count data from the Sloan Digital Sky Survey is 665pc\ (Chen et al. 2001). However, Chen et al.\ (2001) insist on the difficulty to converge to a definitive answer: some other studies indicate that the scale height could be of the order of 1 kpc\ (Gilmore 1984; Ojha et al. 1996). We shall reject the potentials that are completely inconsistent with those characteristics ($0.6 \, {\rm kpc}<h_z<1 \, {\rm kpc}$) and in particular those with a thick disc more massive than the thin disc.  

In order to determine the interval of axis ratios that should be considered for the thick disc, we have fitted the vertical mass distribution corresponding to a simple Kuzmin-Kutuzov potential with $r_S=1$ to an exponential law $e^{-z/h_z}$. We conclude that the potentials with $1.3 \le \epsilon_{\rm thick} \le 2$ have a scale height between 665 pc and 1 kpc and are the ones we should examine in detail. For each axis ratio, Table 1 gives the corresponding scale height. However, Figure 2 reveals that the exponential fit is not valid for the axis ratios used to model the thin disc, which is not surprising since a thin disc could be better understood as a superposition of isothermal sheets than by a simple exponential law.

\begin{figure*}
\vbox{\hspace{-8.8cm}
\psfig{figure=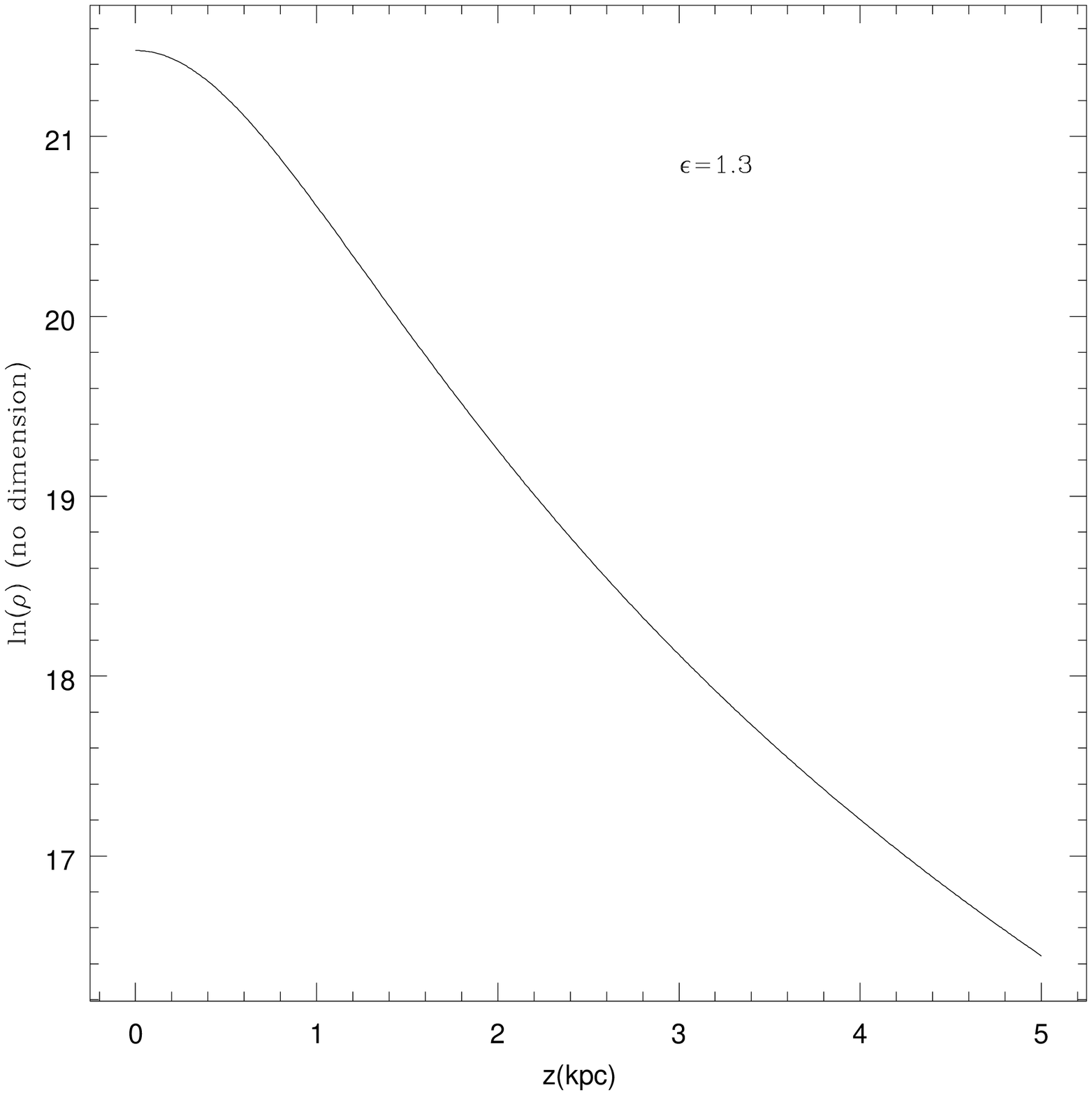,width=8.8cm,angle=0}}

\vbox{\vspace{-8.8cm}\hspace{8.8cm}
\psfig{figure=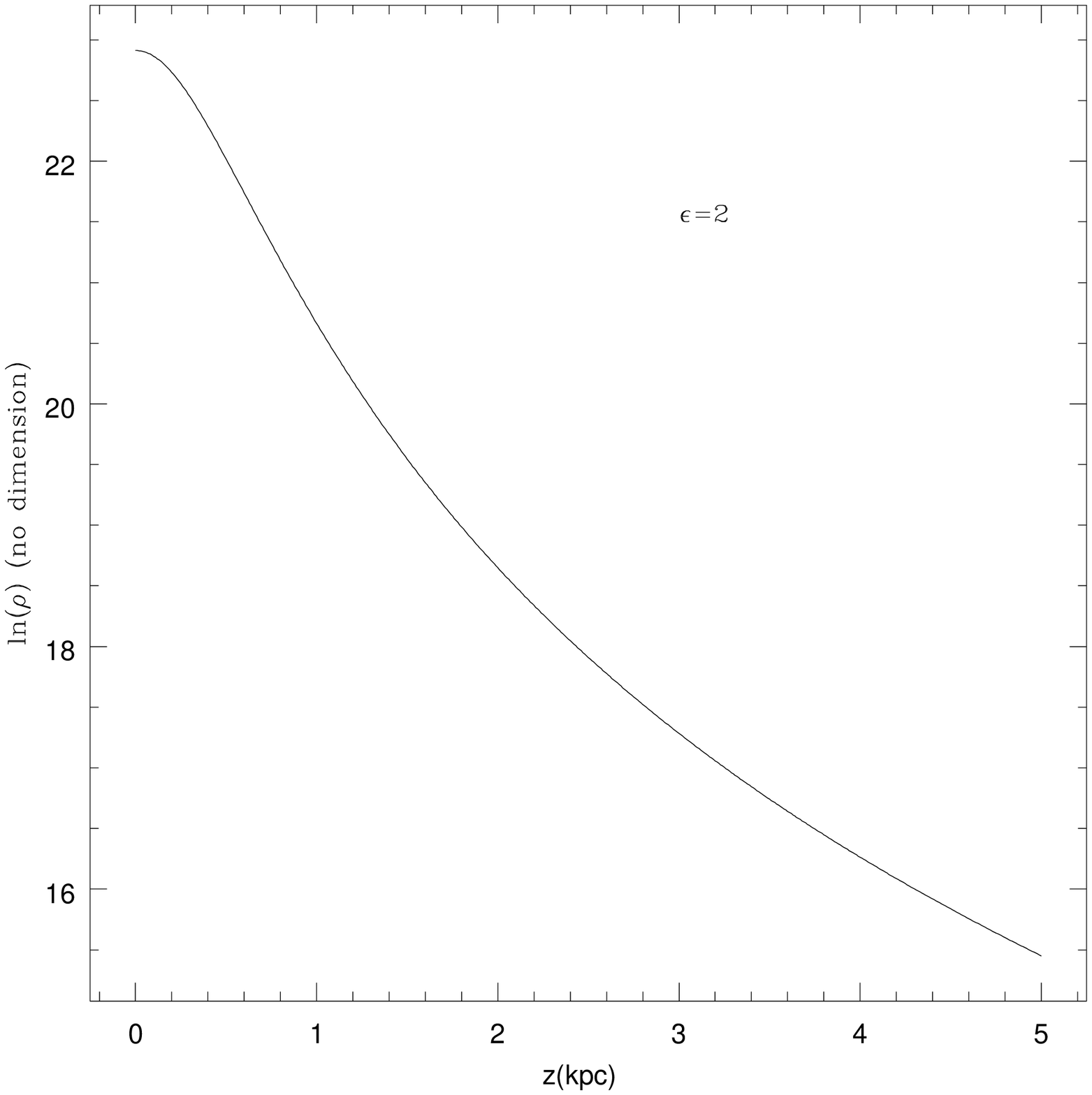,width=8.8cm,angle=0}}

\vbox{\hspace{-8.8cm}
\psfig{figure=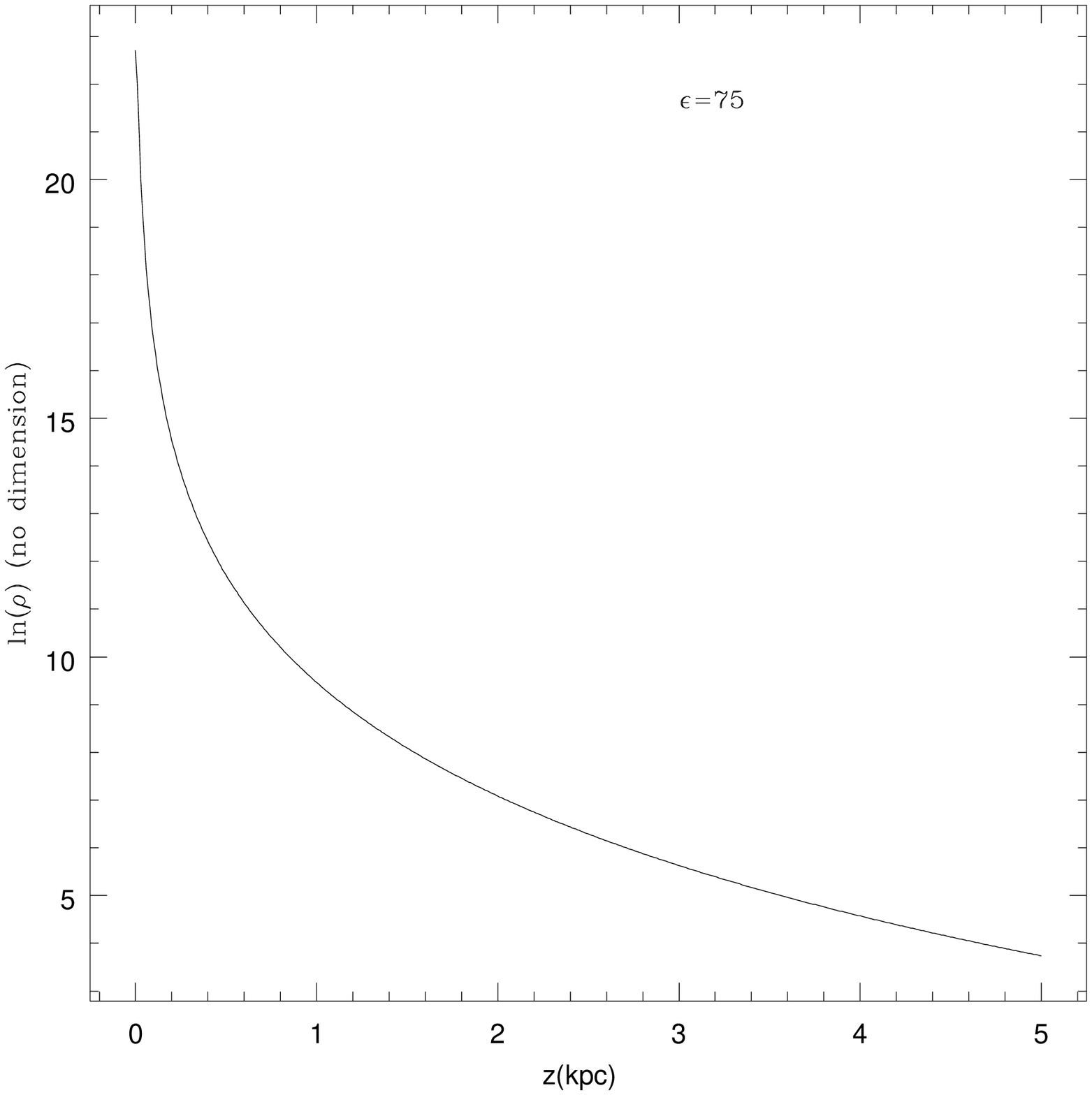,width=8.8cm,angle=0}}

\vbox{\vspace{-8.8cm}\hspace{8.8cm}
\psfig{figure=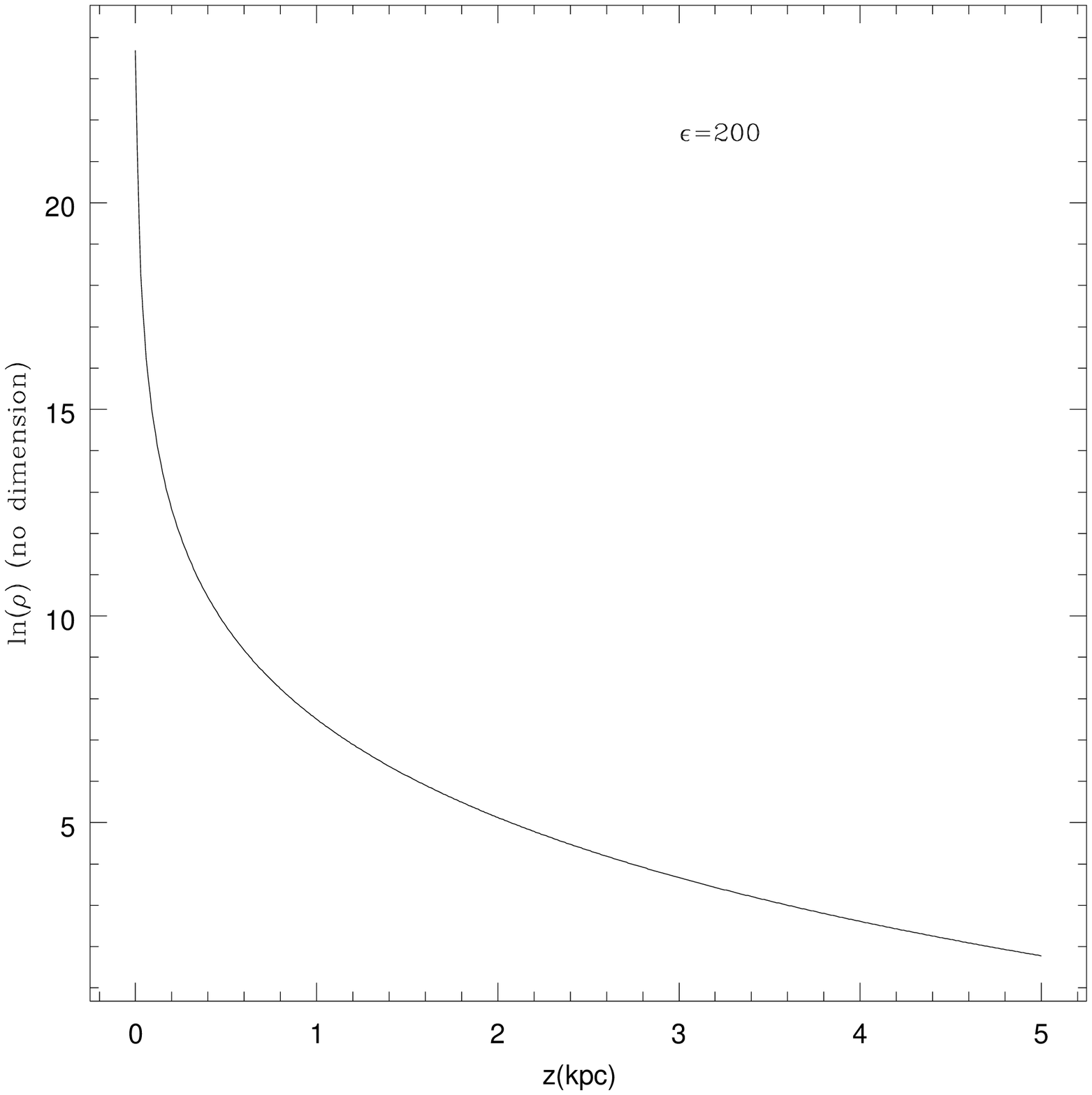,width=8.8cm,angle=0}}

\caption{Profile of the logarithm of vertical density at $\varpi=8$ kpc for Kuzmin-Kutuzov potentials with $r_S=1$ and $\epsilon=1.3$ (top left), $\epsilon=2$ (top right), $\epsilon=75$ (bottom left) and $\epsilon=200$ (bottom right). Only the two first cases resemble exponentials.}
\end{figure*}

\begin{table}
\caption[]{Column 1 contains the axis ratio of the coordinate surfaces for the thick disc. Column 2 gives the corresponding scale height for a Kuzmin-Kutuzov potential with $r_S=1$.}
\begin{tabular}{l | l}
\hline
$\epsilon$ & $h_z$(pc) \\
\hline
1.3 & 914 \\
1.4 & 834 \\
1.5 & 782 \\
1.8 & 696 \\
2 & 663 \\
\hline
\end{tabular}
\end{table}

\subsection{The final selection}

\begin{table*}
\caption[]{The characteristics of the two-component BD potentials with $\epsilon_{\rm disc}=50, 75, 130, 200$  and $\epsilon_{\rm halo}=1.005, 1.01, 1.02, 1.03$ have been examined. The potentials satisfying the selection criteria are listed in this table (with a step of 0.01 for the relative contribution of the disc). Columns 1 and 2 contain the axis ratios of the coordinate surfaces for the disc and the halo. Column 3 contains the relative contribution of the disc to the total mass. Column 4 contains the extent of the flat part of the rotation curve (see Eq. \ref{eq:ef}). Column 5 contains the minimum and maximum local spatial density, while column 6 contains the minimal and maximal local radial derivative of the circular velocity, each time for the two extreme positions of the sun.}
\begin{tabular}{l | l | l | l | c | c}
\hline
$\epsilon_{\rm disc}$ & $\epsilon_{\rm halo}$ & $k_{\rm disc}$ & $E_F$ & $\rho_\odot$ in $M_\odot {\rm pc}^{-3}$ & $\frac{{\rm d}v_c}{{\rm d}\varpi}(\varpi_\odot)$ in ${\rm km}\,{\rm s}^{-1}{\rm kpc}^{-1}$ \\
\hline
75 & 1.02 & 0.11 & 13.07 & 0.04, 0.06 & $-1.90$, $-1.18$ \\
75 & 1.02 & 0.12 & 11.74 & 0.04, 0.06 & $-2.03$, $-1.42$ \\
75 & 1.02 & 0.13 & 10.43 & 0.04, 0.07 & $-2.25$, $-1.74$ \\
75 & 1.02 & 0.14 & 9.25 & 0.04, 0.07 & $-2.46$, $-2.05$ \\
75 & 1.02 & 0.15 & 8.20 & 0.05, 0.08 & $-2.68$, $-2.37$ \\
130 & 1.02 & 0.08 & 17.16 & 0.04, 0.06 & $-2.07$,$-1.12$ \\
130 & 1.02 & 0.09 & 16.03 & 0.04, 0.07 & $-1.75$, $-0.85$ \\
130 & 1.02 & 0.10 & 14.56 & 0.05, 0.08 & $-1.73$, $-0.92$ \\
130 & 1.02 & 0.11 & 13.03 & 0.05, 0.09 & $-1.86$, $-1.15$ \\
130 & 1.02 & 0.12 & 11.70 & 0.06, 0.10 & $-2.00$, $-1.39$ \\
130 & 1.02 & 0.13 & 10.38 & 0.06, 0.11 & $-2.22$, $-1.72$ \\
130 & 1.02 & 0.14 & 9.20 & 0.07, 0.11 & $-2.44$, $-2.04$ \\
130 & 1.02 & 0.15 & 8.08 & 0.07, 0.12 & $-2.69$, $-2.38$ \\
200 & 1.02 & 0.08 & 17.21 & 0.04, 0.09 & $-2.02$, $-1.07$ \\
200 & 1.02 & 0.09 & 16.07 & 0.05, 0.10 & $-1.71$, $-0.81$ \\
200 & 1.02 & 0.10 & 14.59 & 0.06, 0.12 & $-1.69$, $-0.88$ \\
200 & 1.02 & 0.11 & 13.05 & 0.07, 0.13 & $-1.82$, $-1.12$ \\
200 & 1.02 & 0.12 & 11.64 & 0.07, 0.14 & $-2.00$, $-1.40$ \\
200 & 1.02 & 0.13 & 10.32 & 0.08, 0.16 & $-2.22$, $-1.72$ \\
200 & 1.02 & 0.14 & 9.14 & 0.08, 0.17 & $-2.44$, $-2.04$ \\
200 & 1.02 & 0.15 & 8.08 & 0.09, 0.18 & $-2.66$, $-2.36$ \\
\hline
\end{tabular}
\end{table*}

\begin{table*}
\caption[]{The three-component potentials with $\epsilon_{\rm thin}=200$ and $\epsilon_{\rm halo}=1.01$ that satisfy the selection criteria are listed in this table (with a step of 0.01 for the relative contribution of the two discs). Columns 1, 2, 3 contain the axis ratios of the coordinate surfaces for the two discs and the halo. Columns 4 and 5 contain the relative contribution of the repectively thin and thick disc to the total mass. Column 6 contains the extent of the flat part of the rotation curve (see Eq. \ref{eq:ef}). Column 7 contains the minimum and maximum local spatial density, while column 8 contains the minimal and maximal local radial derivative of the circular velocity, each time for the two extreme positions of the sun.}
\begin{tabular}{l | l | l | l | l | l | c | c}
\hline
$\epsilon_{\rm thin}$ & $\epsilon_{\rm thick}$ & $\epsilon_{\rm halo}$ & $k_{\rm thin}$ & $k_{\rm thick}$ & $E_F$ & $\rho_\odot$ in $M_\odot {\rm pc}^{-3}$ & $\frac{{\rm d}v_c}{{\rm d}\varpi}(\varpi_\odot)$ in ${\rm km}\,{\rm s}^{-1}{\rm kpc}^{-1}$\\
\hline
200 & 1.3 & 1.01 & 0.09 & 0.07 & 10.69 & 0.06, 0.12 & $-1.34$, $-1.13$\\
200 & 1.3 & 1.01 & 0.09 & 0.08 & 10.06 & 0.06, 0.12 & $-1.64$, $-1.45$\\
200 & 1.3 & 1.01 & 0.10 & 0.06 & 9.51 & 0.07, 0.15 & $-1.35$, $-1.29$\\
200 & 1.3 & 1.01 & 0.10 & 0.07 & 8.95 & 0.07, 0.14 & $-1.64$, $-1.59$\\
200 & 1.3 & 1.01 & 0.10 & 0.08 & 8.45 & 0.07, 0.13 & $-1.92$, $-1.88$\\
200 & 1.3 & 1.01 & 0.11 & 0.04 & 8.51 & 0.09, 0.18 & $-1.22$, $-1.17$\\
200 & 1.3 & 1.01 & 0.11 & 0.05 & 8.02 & 0.08, 0.17 & $-1.51$, $-1.45$\\
200 & 1.4 & 1.01 & 0.09 & 0.07 & 9.81 & 0.06, 0.13 & $-1.46$, $-1.32$\\
200 & 1.4 & 1.01 & 0.09 & 0.08 & 9.11 & 0.06, 0.12 & $-1.77$, $-1.65$\\
200 & 1.4 & 1.01 & 0.10 & 0.06 & 8.63 & 0.07, 0.14 & $-1.50$, $-1.47$\\
200 & 1.4 & 1.01 & 0.10 & 0.07 & 8.07 & 0.07, 0.14 & $-1.78$, $-1.78$\\
200 & 1.5 & 1.01 & 0.08 & 0.07 & 10.86 & 0.06, 0.11 & $-1.25$, $-1.00$\\
200 & 1.5 & 1.01 & 0.09 & 0.06 & 9.93 & 0.07, 0.13 & $-1.21$, $-1.09$\\
200 & 1.5 & 1.01 & 0.09 & 0.07 & 9.10 & 0.06, 0.13 & $-1.54$, $-1.45$\\
200 & 1.5 & 1.01 & 0.09 & 0.08 & 8.34 & 0.06, 0.12 & $-1.86$, $-1.80$\\
200 & 1.5 & 1.01 & 0.10 & 0.06 & 8.01 & 0.07, 0.15 & $-1.59$, $-1.57$\\
200 & 1.8 & 1.01 & 0.08 & 0.07 & 9.42 & 0.06, 0.12 & $-1.34$, $-1.22$\\
200 & 1.8 & 1.01 & 0.09 & 0.06 & 8.55 & 0.07, 0.13 & $-1.34$, $-1.32$\\
200 & 2 & 1.01 & 0.08 & 0.07 & 8.75 & 0.06, 0.12 & $-1.38$, $-1.31$\\
\hline

\end{tabular}
\end{table*}

\begin{table*}
\caption[]{The three-component potentials with $\epsilon_{\rm thin}=75$ and $\epsilon_{\rm halo}=1.01$ that satisfy the selection criteria are listed in this table (with a step of 0.01 for the relative contribution of the two discs). Columns have the same meaning as in table 3.}
\begin{tabular}{l | l | l | l | l | l | c | c}
\hline
$\epsilon_{\rm thin}$ & $\epsilon_{\rm thick}$ & $\epsilon_{\rm halo}$ & $k_{\rm thin}$ & $k_{\rm thick}$ & $E_F$ & $\rho_\odot$ in $M_\odot {\rm pc}^{-3}$ & $\frac{{\rm d}v_c}{{\rm d}\varpi}(\varpi_\odot)$ in ${\rm km}\,{\rm s}^{-1}{\rm kpc}^{-1}$\\
\hline
75 & 1.3 & 1.01 & 0.11 & 0.05 & 8.21 & 0.05, 0.07 & $-1.49$, $-1.45$\\
75 & 1.4 & 1.01 & 0.11 & 0.04 & 8.06 & 0.05, 0.07 & $-1.32$, $-1.27$\\
75 & 1.5 & 1.01 & 0.10 & 0.05 & 8.87 & 0.04, 0.07 & $-1.26$, $-1.23$\\
\hline

\end{tabular}
\end{table*}

\begin{table*}
\caption[]{Among the class of potentials defined in section 3, five different potentials regarding form and features have been selected. Columns 1, 2, 3 contain the axis ratios of the coordinate surfaces for the two discs and the halo. Columns 4 and 5 contain the relative contribution of the repectively thin and thick disc to the total mass. Column 6 contains the extent of the flat part of the rotation curve (see Eq. \ref{eq:ef}). Column 7 contains the scale factor which corresponds to the focal distance of the coordinate system of the dimensional potential. Column 8 contains the minimum and maximum local spatial density in $M_\odot {\rm pc}^{-3}$, while column 9 contains the minimum and maximum local radial derivative of the circular velocity in ${\rm km}\,{\rm s}^{-1}{\rm kpc}^{-1}$ and column 10 the minimum and maximum total mass of the Galaxy in $10^{11} M_\odot$, each time for the two extreme positions of the sun. The scale length in the equatorial plane down to 3 kpc has been calculated and is presented in column 11 (in kpc).}
\begin{tabular}{l| l | l | l | l | l | l | l | c | c | c | c}
\hline
  & $\epsilon_{\rm thin}$ & $\epsilon_{\rm thick}$ & $\epsilon_{\rm halo}$ & $k_{\rm thin}$ & $k_{\rm thick}$ & $E_F$ & $r_S$ & $\rho_\odot$ & $\frac{{\rm d}v_c}{{\rm d}\varpi}(\varpi_\odot)$ & $M$ & scale length \\
\hline
I & 75 & 1.5 & 1.02 & 0.13 & 0.01 & 9.86 & 0.93 & 0.04, 0.07 & $-2.51$, $-2.05$ & 2.37, 2.41 & 2.73 \\
II & 200 & 1.8 & 1.02 & 0.10 & 0.01 & 13.44 & 0.88 & 0.06, 0.12 & $-2.05$, $-1.31$ & 2.37, 2.41 & 2.63 \\
III & 200 & 1.3 & 1.01 & 0.11 & 0.04 & 8.51 & 0.95 & 0.09, 0.18 & $-1.22$, $-1.17$ & 3.19, 3.22 & 2.65\\
IV & 75 & 1.8 & 1.01 & 0.11 & 0.015 & 9.07 & 0.98 & 0.05, 0.08 & $-0.62$, $-0.55$ & 3.56, 3.58 & 2.78\\
V & 200 & 1.3 & 1.005 & 0.07 & 0.01 & 18.30 & 1.01 & 0.11, 0.23 & $+0.69$, $+1.47$ & 6.13, 6.20 & 2.72\\
\hline

\end{tabular}
\end{table*}

\begin{figure*}
\vbox{\hspace{-8.8cm}
\psfig{figure=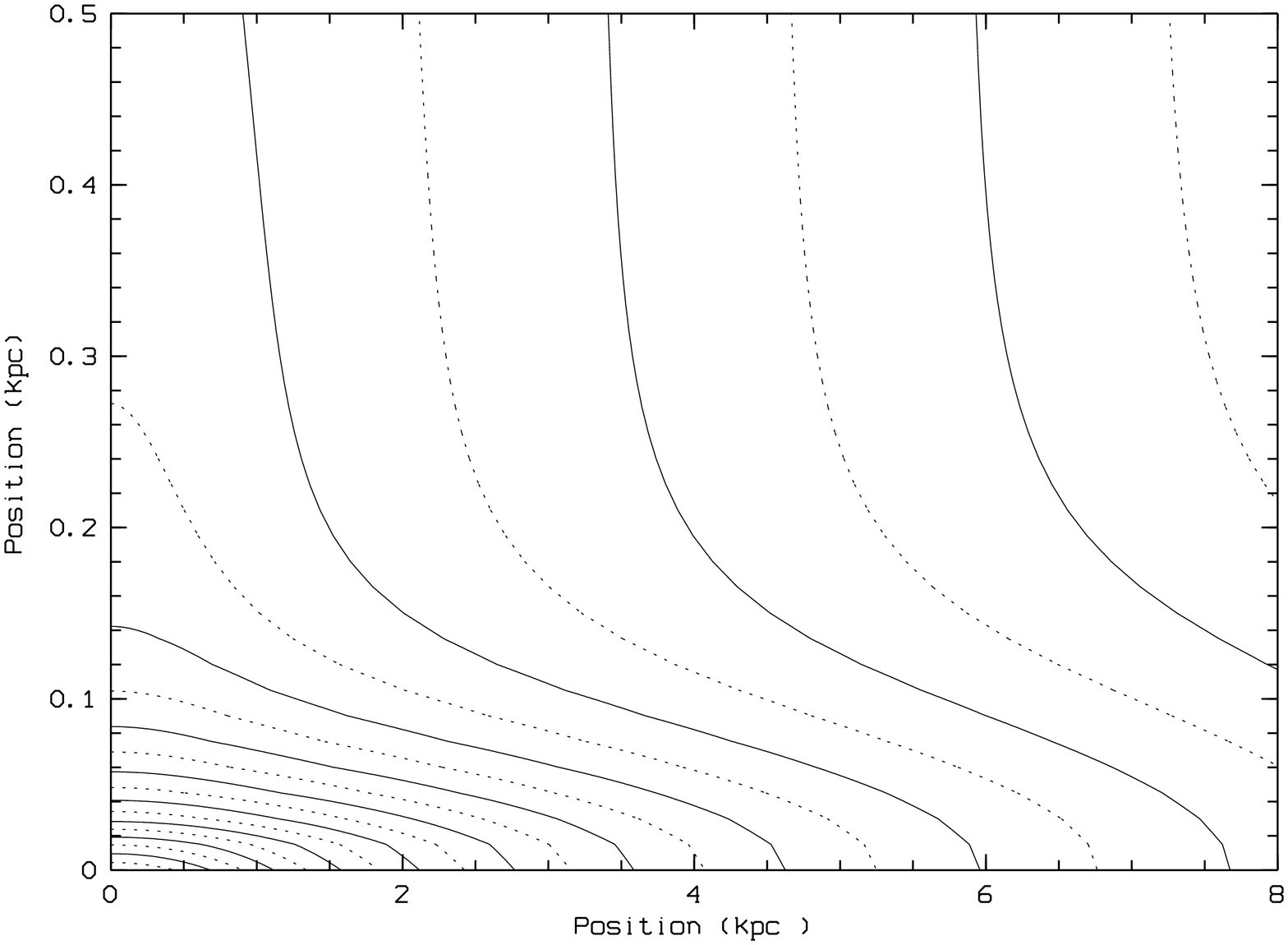,width=4cm,angle=-90}}

\vbox{\vspace{-8.8cm}\hspace{8.8cm}
\psfig{figure=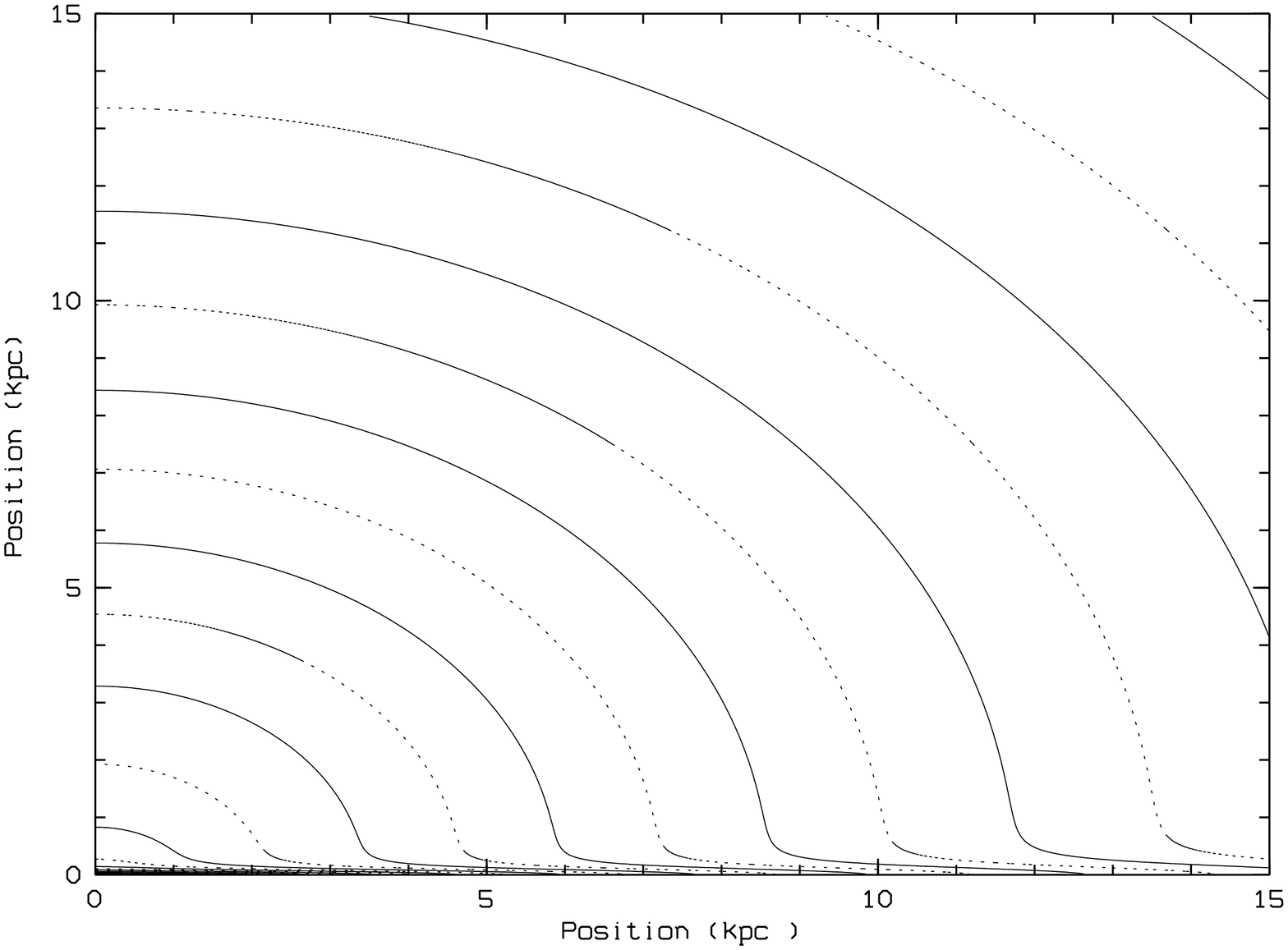,width=4cm,angle=-90}}

\vbox{\hspace{-8.8cm}
\psfig{figure=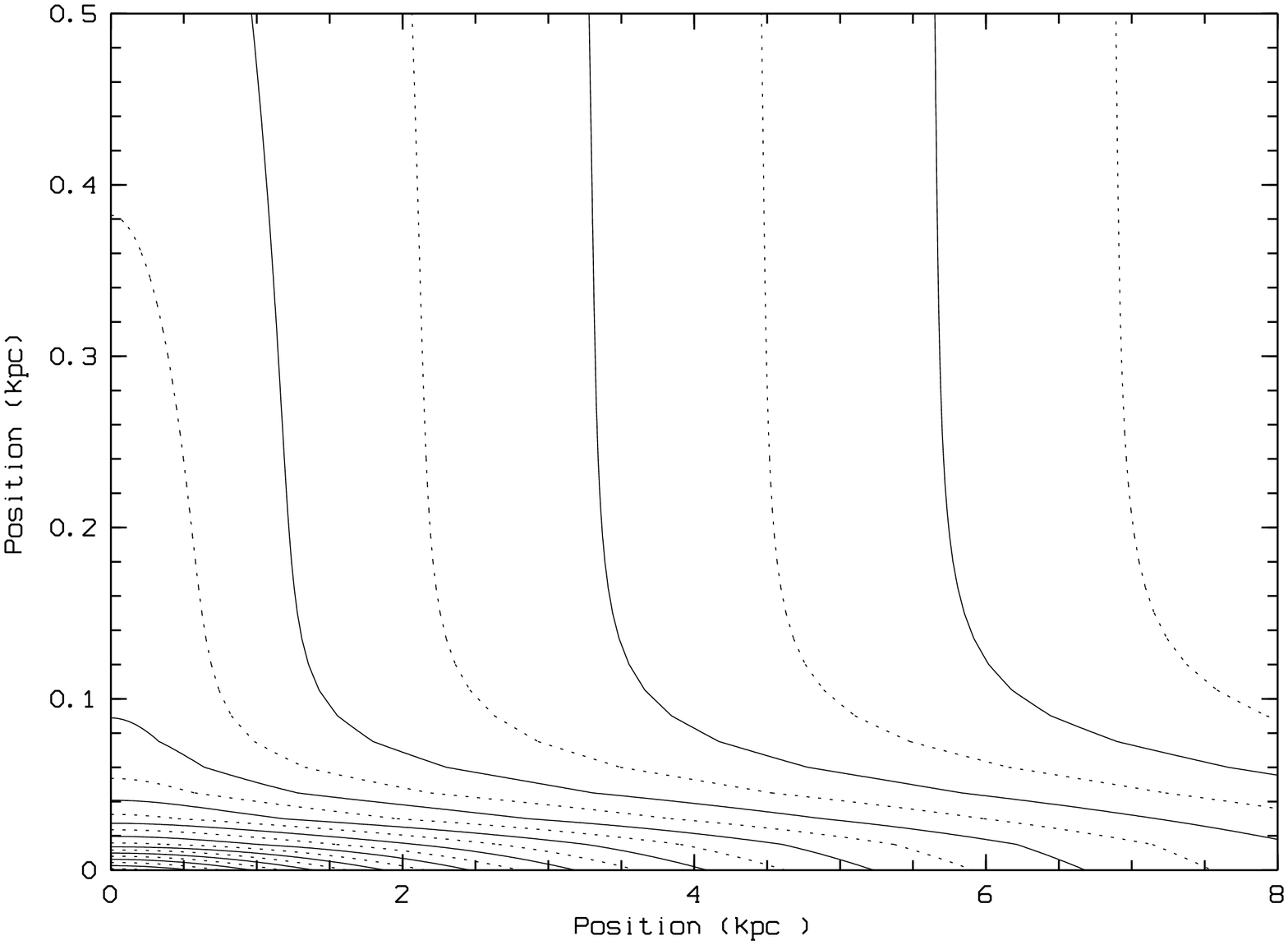,width=4cm,angle=-90}}

\vbox{\vspace{-8.8cm}\hspace{8.8cm}
\psfig{figure=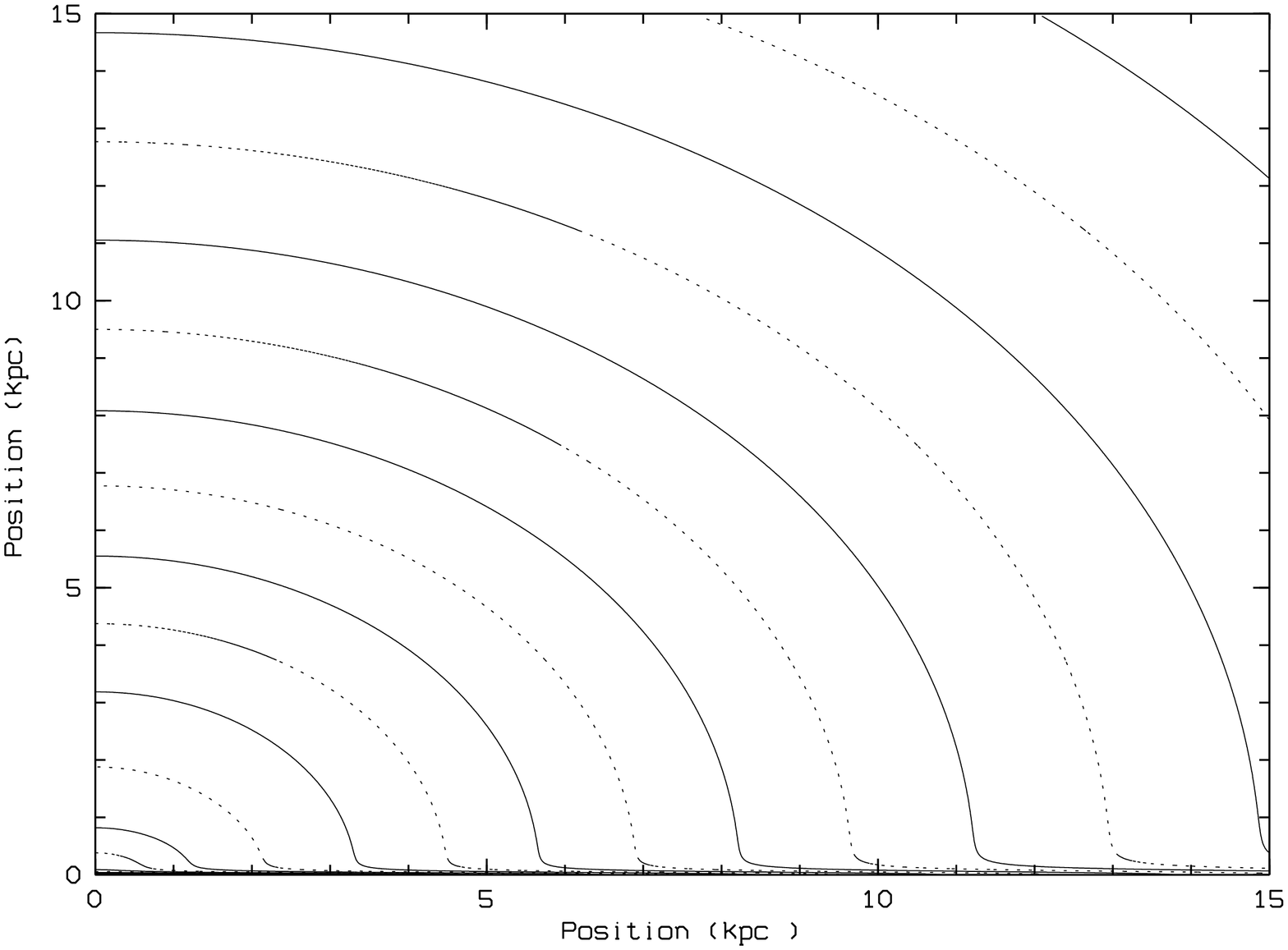,width=4cm,angle=-90}}

\vbox{\hspace{-8.8cm}
\psfig{figure=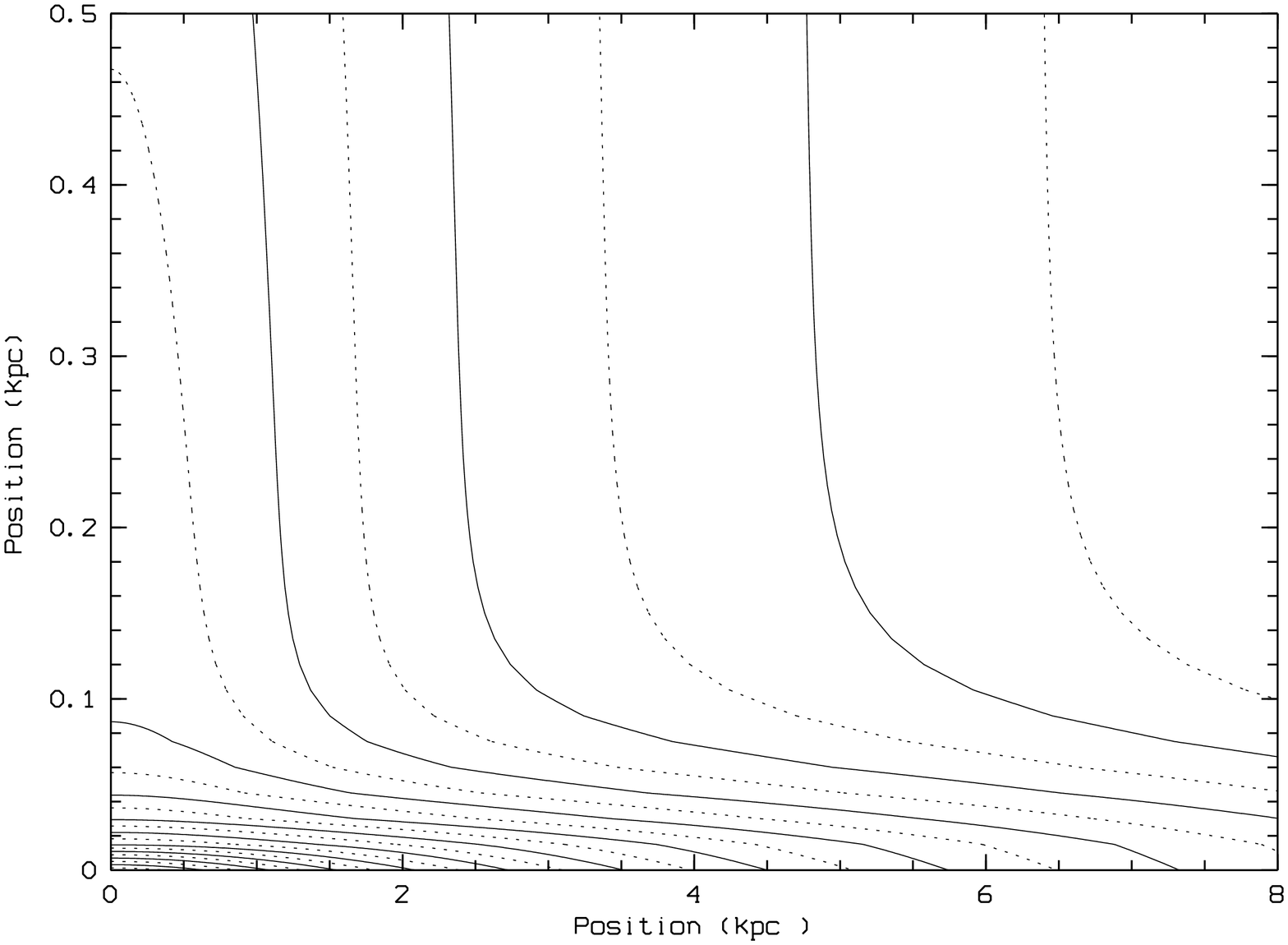,width=4cm,angle=-90}}

\vbox{\vspace{-8.8cm}\hspace{8.8cm}
\psfig{figure=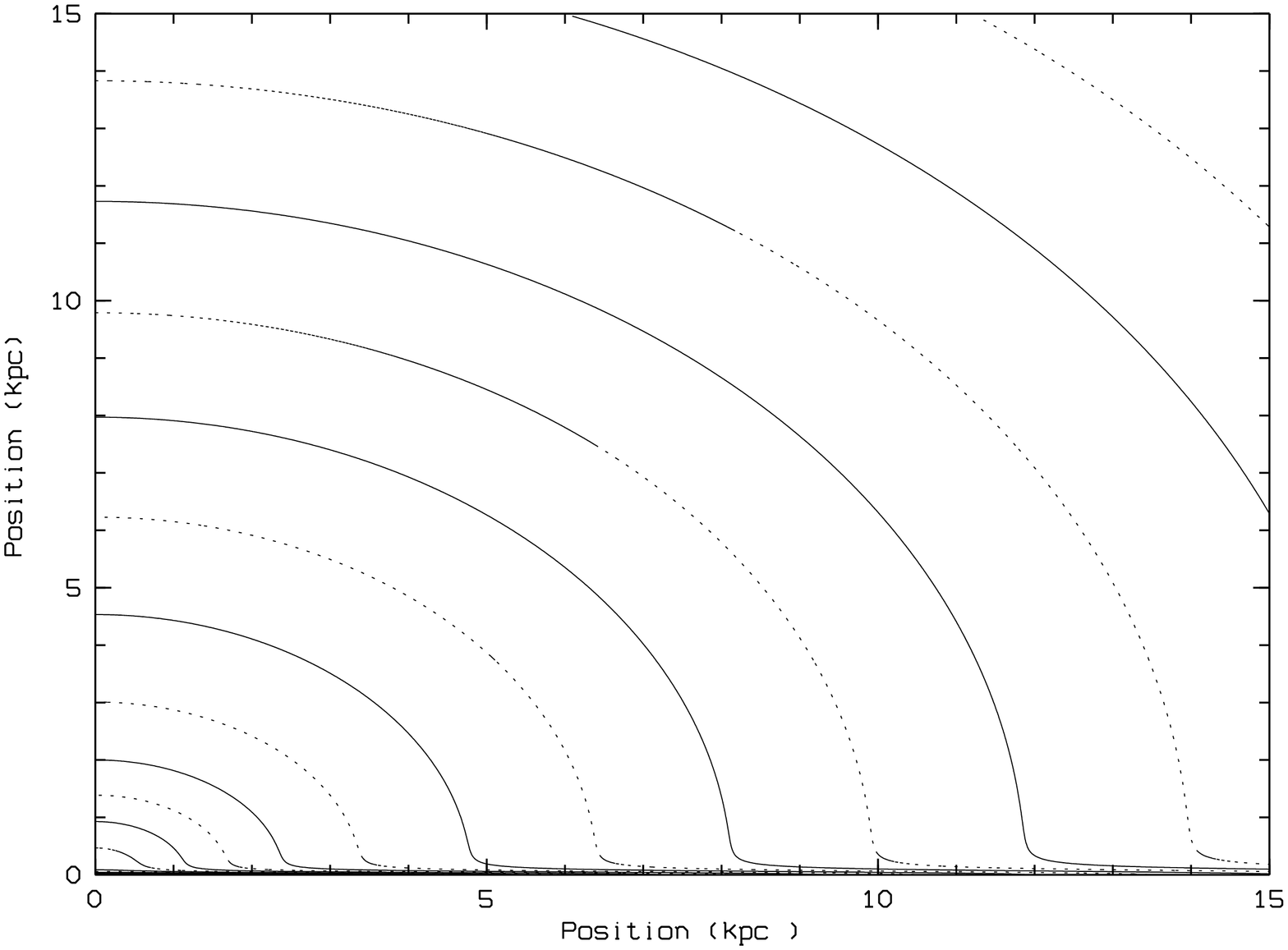,width=4cm,angle=-90}}

\vbox{\hspace{-8.8cm}
\psfig{figure=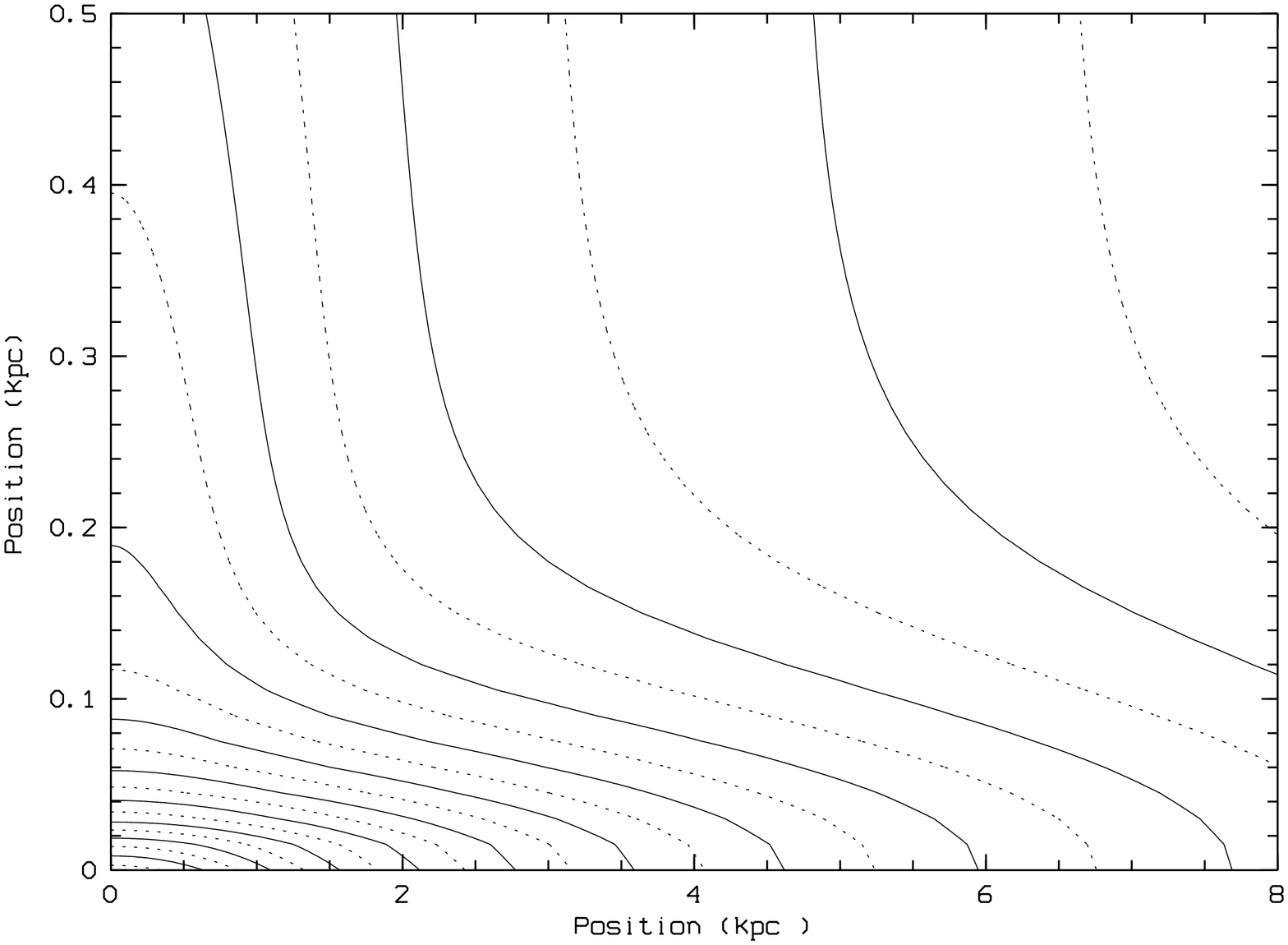,width=4cm,angle=-90}}

\vbox{\vspace{-8.8cm}\hspace{8.8cm}
\psfig{figure=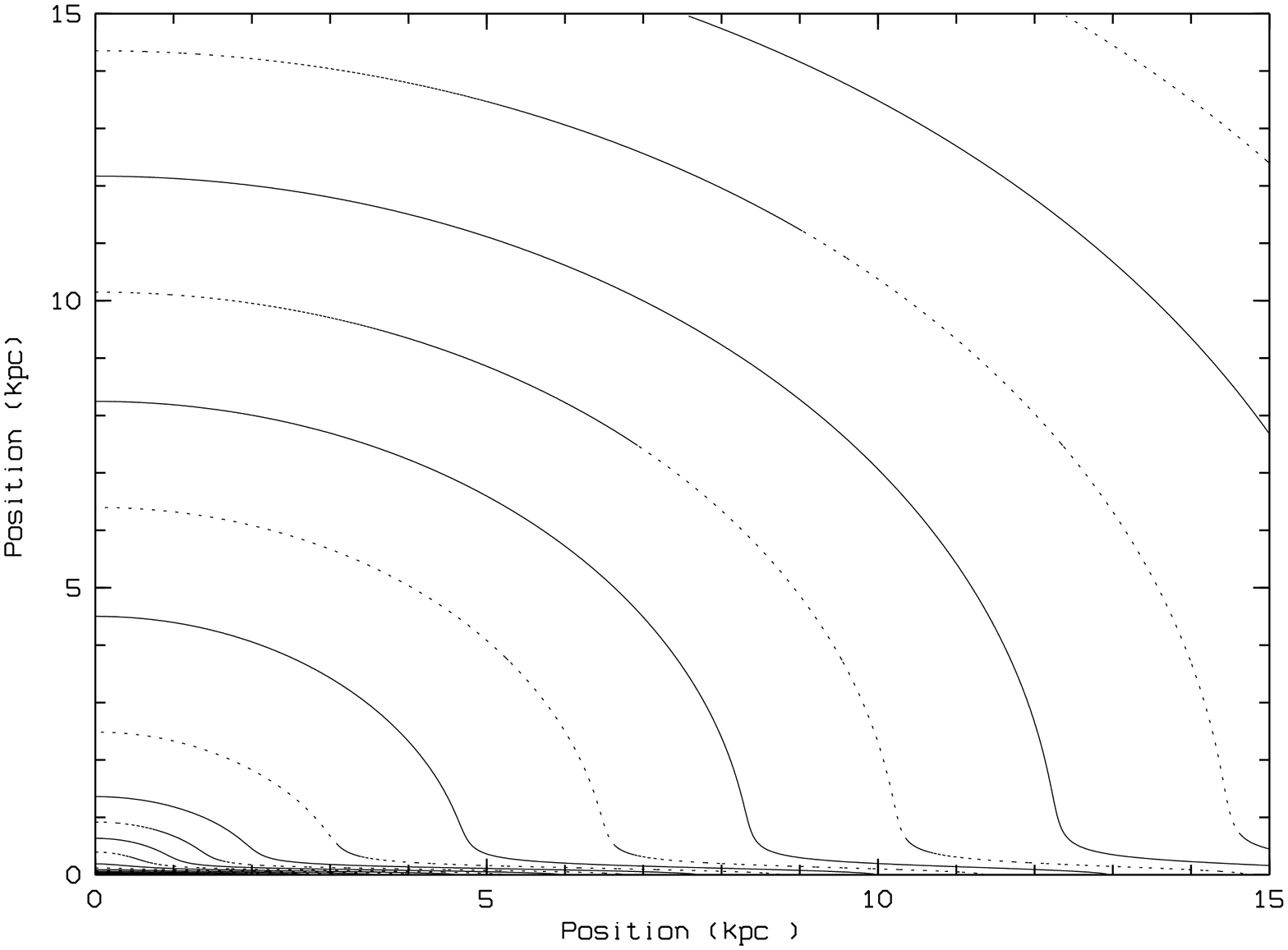,width=4cm,angle=-90}}

\vbox{\hspace{-8.8cm}
\psfig{figure=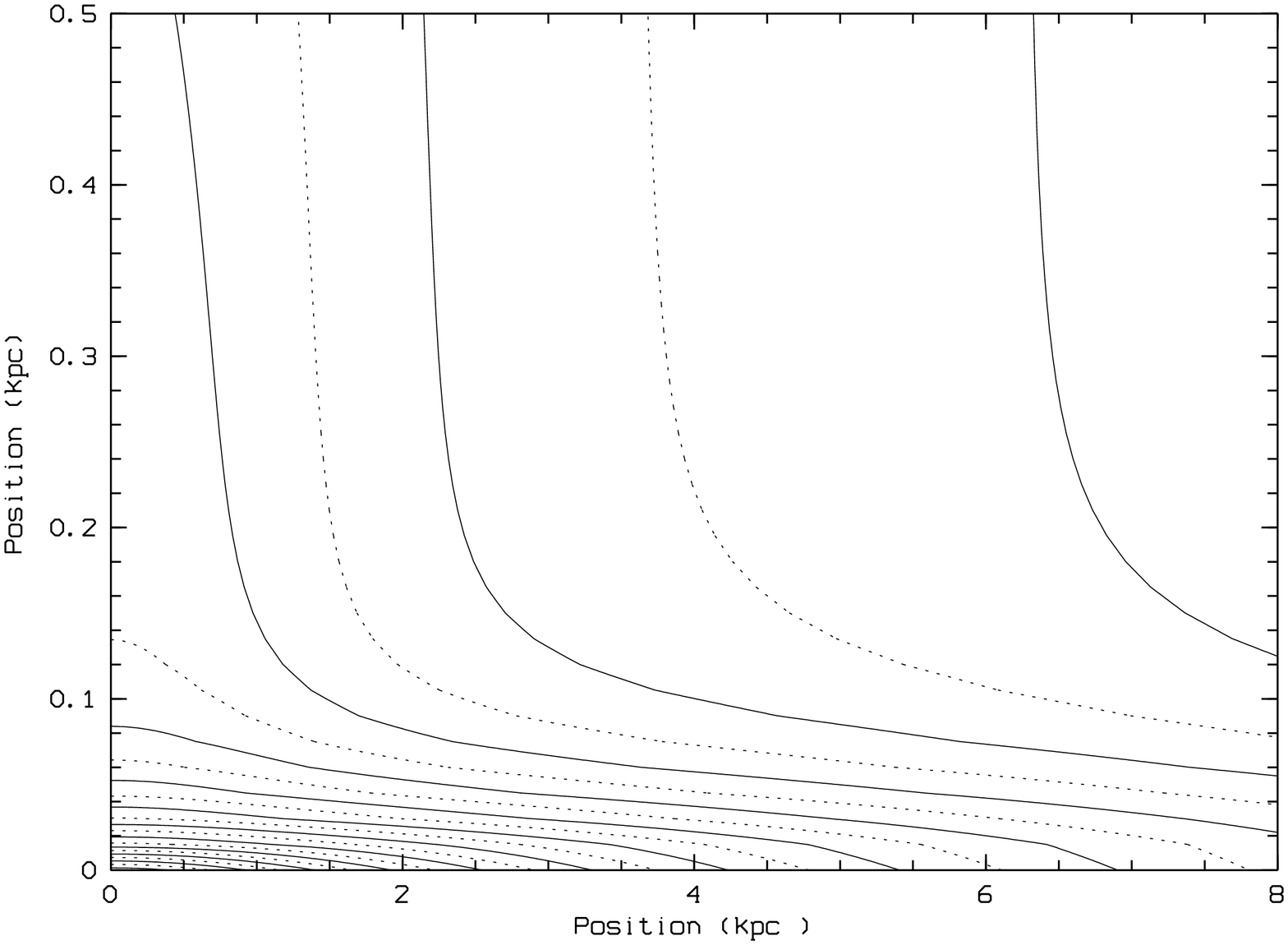,width=4cm,angle=-90}}

\vbox{\vspace{-8.8cm}\hspace{8.8cm}
\psfig{figure=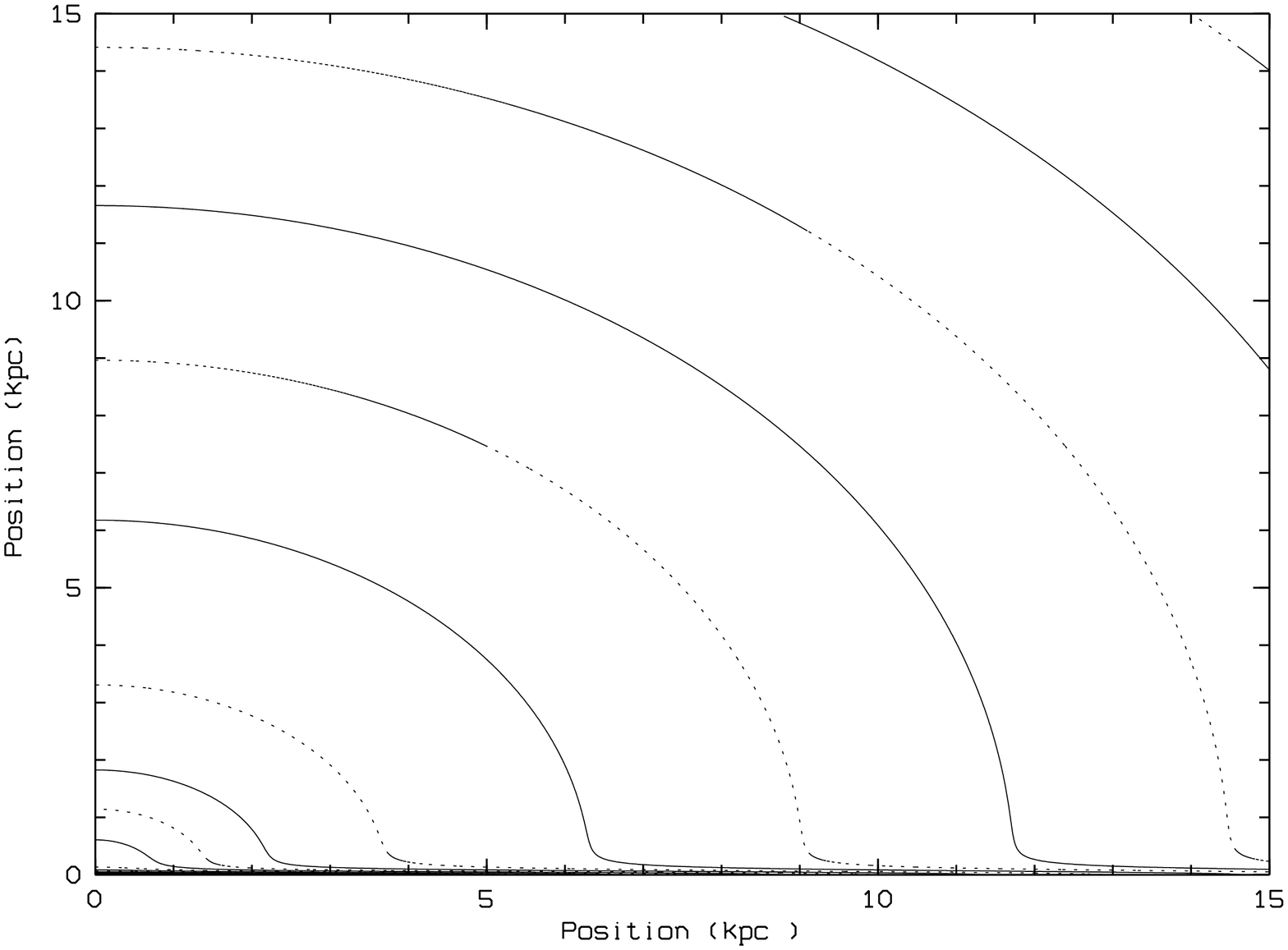,width=4cm,angle=-90}}

\caption{Mass isodensity curves in a meridional plane for the five potentials of Table 5, with two different scales (left panel: zoom on the disc, right panel: large scale view of the halo).}
\end{figure*}

\begin{figure*}
\psfig{figure=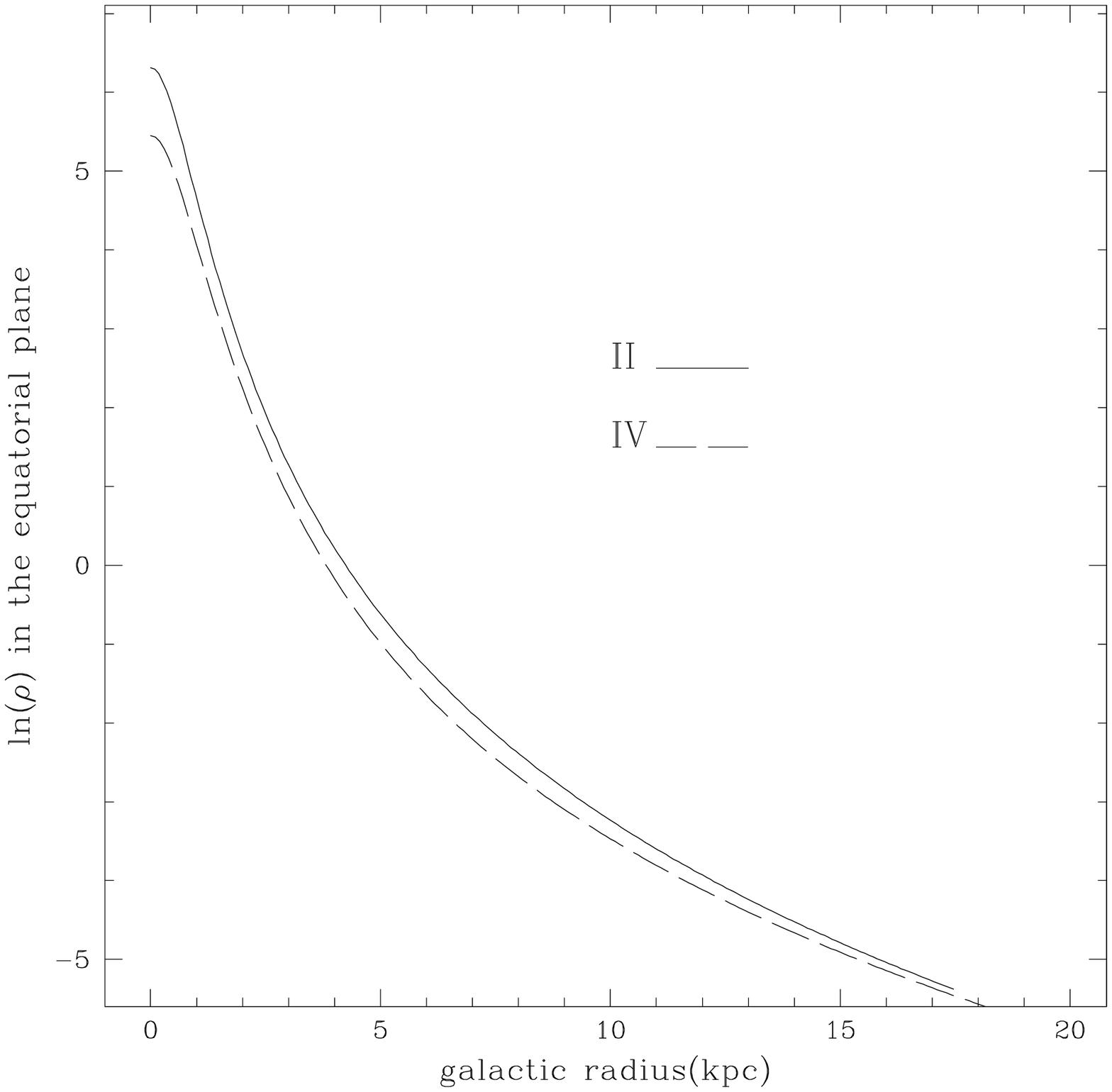,width=12.cm,angle=0} \caption{The logarithm of the mass density in the equatorial plane for the two potentials with extreme scale lengths.These curves very much resemble each other, and the effective bulge appears clearly.}
\end{figure*}

\begin{figure*}
\psfig{figure=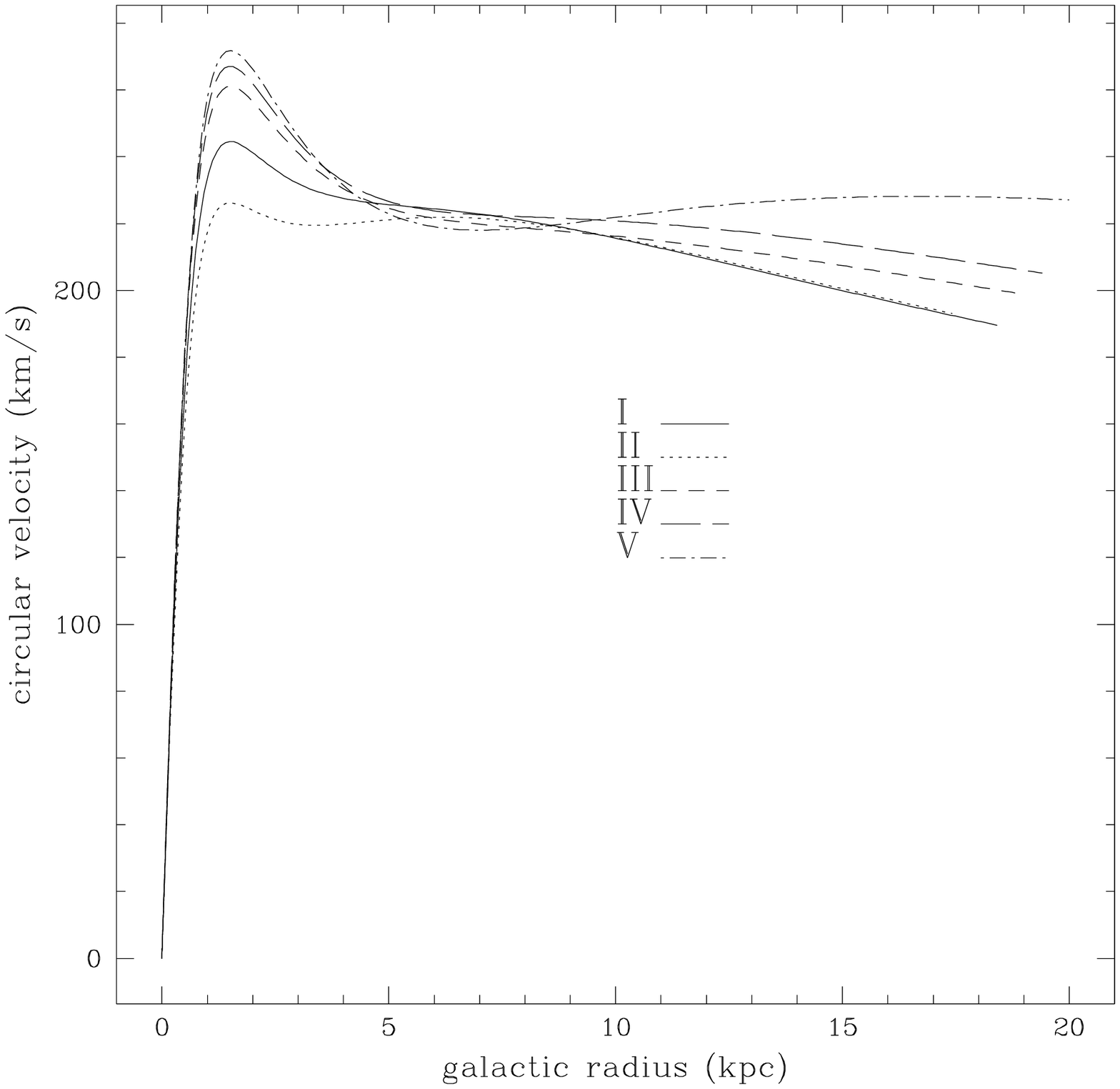,width=12.cm,angle=0} \caption{The rotation curves of the five selected potentials of Table 5. The total mass used to plot thes curves is the mean total mass of the two extreme values of Table 5.}
\end{figure*}

In this section, we look for some three-component potentials with different forms and features, all satisfying the selection criteria defined in section 3.3 and consistent with what is known about the thick disc.

First of all, we look for the two-component BD potentials that satisfy the new selection criteria: they are listed in Table 2. All the two-component potentials with $\epsilon_{\rm disc}=50, 75, 130, 200$  and $\epsilon_{\rm halo}=1.005, 1.01, 1.02, 1.03$ are examined. We do not consider discs with $a/c > 200$ because then the uncertainty on $\rho_\odot$ becomes too large. We see in Table 2 that, in order to reproduce the Oort constants in the two-component framework, the shape of the halo cannot vary ($\epsilon_{\rm halo}=1.02$).

If we take the two-component potentials as a starting point, there are two ways to add a thick disc. The first way is to decrease the contribution of the halo and to put the remaining mass into the thick disc: the local density in the solar neighbourhood is then slightly larger while the rotation curve is decreasing faster. If we take the third potential of Table 2 as a starting point, the first potential of Table 5 (potential I) illustrates this first case. The other way is to decrease the contribution of the thin disc and to put the remaining mass into the thick disc: the local density is then slightly decreasing while the rotation curve is more flat. If we take the fourth potential of Table 2 as a starting point, the first potential of Table 5 illustrates this second case: this potential (potential I) is a three-component potential satisfying the selection criteria and is selected to be analyzed in detail and confronted with kinematical surveys in the future.

The presence of a third component allows more freedom for the shape of the halo, so we look for three-component potentials with a halo rounder than $\epsilon_{\rm halo}=1.02$. In order to keep the rotation curve flat and retain the local density as well as the Oort constants in the allowed interval, we need to couple a very thin disc with the rounder halo: indeed, our investigations show that no solution can be found for $\epsilon_{\rm thin}=50$ and $\epsilon_{\rm halo}=1.01$. However, if we take $\epsilon_{\rm thin}=200$, Table 3 gives solutions for a halo with $\epsilon_{\rm halo}=1.01$: we select the solution where the mass of the thick disc relative to the thin disc is the smallest (potential III of Table 5). Remark that a similar Table for $\epsilon_{\rm halo}=1.02$ would contain 292 entries and is omitted here. There are much less solutions when $\epsilon_{\rm thin}=75$, as can be seen in Table 4. However, as stated in section 3.3, we do not assign high priority to the Oort constants, and we select a potential  with  $\epsilon_{\rm thin}=75$, $\epsilon_{\rm halo}=1.01$ and a relative mass of the thick disc relative to the thin disc of 13\% (i.e. a smaller fraction than any of the solutions of Table 4), but with a quite large local radial derivative of the circular speed (potential IV of Table 5).

For $\epsilon_{\rm halo}=1.005$, it is totally impossible to find a potential satisfying the Oort constants criterion: the radial derivative of the circular speed in the solar neighbourhood is always positive. Nevertheless, if one is willing to ignore the estimates of $A$ and $B$, one could select a potential with $\epsilon_{\rm halo}=1.005$ and reasonably low values for the cirular speed radial derivative in the solar neighbourhood (potential V of Table 5).

Finally, we select a potential (potential II) satisfying all the criteria, for which the interval in $\rho_\odot$ is precisely $\lbrack0.06 M_\odot {\rm pc}^{-3}, 0.12 M_\odot {\rm pc}^{-3}\rbrack$, and which is close to the Chen et al.\ (2001) findings , i.e. $\epsilon_{\rm thin}=200$ which is the thinnest thin disc that we consider in order not to have a too large interval for $\rho_\odot$, a relative mass of the thick disc to the thin disc of 10\%, a scale height of the thick disc of 612.5 pc and a relatively large $E_F$ ($E_F=13.44$).

Table 5 summarizes the main features of the selected St\"ackel potentials with different forms and features and that we shall use for dynamical modeling of the Milky Way: potentials III and V have a very thin disc associated with a quite massive thick disc and a pretty round halo, while potentials I and IV have a thicker thin disc with a quasi-negligible thick disc (Figure 3 shows the mass isodensity curves of each potential in a meridional plane for two different scales). It should be noted that the total masses associated with those potentials are very different and become larger with a rounder halo and that a rounder halo implies that this halo is much more extended. A closer look to the mass density in the equatorial plane indicates that, for each potential, the density grows faster than an exponential in the central 3 kpc corresponding to the bulge region: the potentials have thus an effective bulge, which did enable us to avoid the introduction of an explicit bulge component. We have fitted the mass density in the plane to an exponential law down to $\varpi=3$ kpc in order to check that the scale length of the disc is realistic: the last column of Table 5 gives the scale length corresponding to each selected potential and we conclude that they are realistic but do not distinguish the different potentials. For the potentials with the biggest and smallest scale length, Figure 4 illustrates the shape of the logarithm of the density in the plane (the other potentials have a similar shape for that curve). Finally, Figure 5 shows the rotation curve associated with each of the five selected potentials: the rotation curve of potential II is more flat than the one of potential I in the vicinity of the sun, while the rotation curves of the potentials with $\epsilon_{\rm halo}=1.01$ (potentials III and IV) are even more flat and the one of potential V is slightly increasing.

\section{Conclusions}

In this paper, we have shown that some different simple St\"ackel potentials can fit most known parameters of the Milky Way (especially Hipparcos latest findings). First, we have reviewed the galactic fundamental parameters that any Milky Way potential must match and that investigations following the Hipparcos mission have readjusted. Then, we have generalized the two-component BD potentials by adding a thick disc, and we have studied how the parameters can vary in order to satisfy selection criteria based on the latest observational constraints. We have shown that the presence of a thick disc allows more flexibility in the form of the potentials, especially for the shape of the halo and we have selected five different valid potentials listed in Table 5. It should be noted that, in fact, there could be two different thick discs in the Galaxy, a very thick one (Chiba \& Beers 2000; Gilmore, Wyse \& Norris 2002) and a smaller one (Soubiran, Bienaym\'e \& Siebert 2002): in that case, the three-component modelling presented in this paper could be easily extended, but this would imply a growth of parameter space.

A major result of this paper is that, even though St\"ackel potentials are negligible in the set of all potentials, many of them are still able to match the most recent estimates for the parameters of the Milky Way, and furthermore very simple ones (superpositions of three Kuzmin-Kutuzov potentials) are sufficient to do this.

The potentials defined in this paper have already been used in Famaey et al.\ (2002). We plan to further validate each of the five proposed potentials by confronting them with kinematical surveys. This we shall do by constructing three-integral analytic distribution functions in those potentials, using a quadratic programming technique\ (Dejonghe 1989; for an overview see Dejonghe et al. 2001).

\section*{Acknowledgements}

We thank Dr Alain Jorissen for his permanent assistance, many interesting discussions and some helpful suggestions.

\label{lastpage}

\end{document}